\documentclass[preprint,journal]{vgtc}                % final (journal style)
\ifpdf%                                % if we use pdflatex
  \pdfoutput=1\relax                   % create PDFs from pdfLaTeX
  \usepackage{graphicx}                % allow us to embed graphics files
  \DeclareGraphicsExtensions{.pdf,.png,.jpg,.jpeg} % for pdflatex we expect .pdf, .png, or .jpg files
\else%                                 % else we use pure latex
  \usepackage{graphicx}                % allow us to embed graphics files
  \DeclareGraphicsExtensions{.eps}     % for pure latex we expect eps files
\fi%

\graphicspath{{figures/}{pictures/}{images/}{./}} % where to search for the images

\usepackage{microtype}                 % use micro-typography (slightly more compact, better to read)
\PassOptionsToPackage{warn}{textcomp}  % to address font issues with \textrightarrow
\usepackage{textcomp}                  % use better special symbols
\usepackage{mathptmx}                  % use matching math font
\usepackage{times}                     % we use Times as the main font
\usepackage{cite}                      % needed to automatically sort the references
\usepackage{tabu}                      % only used for the table example
\usepackage{booktabs}                  % only used for the table example
\usepackage{graphicx}
\usepackage{array}
\usepackage{booktabs}
\usepackage{hhline}
\usepackage{multirow}
\newcolumntype{P}[1]{>{\centering\arraybackslash}p{#1}}

\usepackage{soul} % highlight with \hl{}
\usepackage{dirtytalk} % blockquote with \begin{displayquote}
\usepackage{csquotes} % quote with \say{}
\usepackage{censor} % redact with \censor{}
\renewenvironment{quote}%
  {\list{}{\leftmargin=0.12in\rightmargin=0.12in}\item[]}%
  {\endlist}
 
\usepackage{color}
\usepackage[table,xcdraw]{xcolor}

\definecolor{Level1}{rgb}{1,0.84,0.7}
\definecolor{Level1Light}{rgb}{1,0.96,0.88}
\definecolor{Level2}{rgb}{0.58,0.87,0.58}
\definecolor{Level2Light}{rgb}{0.95,1,0.92}
\definecolor{Level3}{rgb}{1,1,0.66}
\definecolor{Level3Light}{rgb}{1,1,0.86}
\definecolor{Level4}{rgb}{0.17,0.51,0.83}
\definecolor{Level4Light}{rgb}{0.93,0.99,1}
\definecolor{Gray}{rgb}{0.9,0.9,0.9}
\definecolor{GrayLight}{rgb}{0.97,0.97,0.97}

\definecolor{DirectQuoteA}{rgb}{0.93,0.92,1}
\definecolor{DirectQuoteB}{rgb}{0.93,0.93,0.93}

\newcommand{\hlc}[2][yellow]{{\sethlcolor{#1}\hl{#2}}}

\newcommand{\hquoteA}[2][black]{\textcolor{black}{\hlc[DirectQuoteA]{\emph{``#2''}}}}
\newcommand{\hquoteB}[2][black]{\textcolor{black}{\hlc[DirectQuoteB]{\emph{``#2''}}}}

\usepackage[nolist,nohyperlinks, smaller]{acronym}
\begin{acronym}
    \acro{SVG}{Scalable Vector Graphics}
    \acro{NL}{natural language}
    \acro{JSON}{JavaScript Object Notation}
    \acro{HTML}{HyperText Markup Language}
    \acro{AI}{Artificial Intelligence}
    \acro{CSS}{Cascading Style Sheets}
    \acro{HTML5}{HTML5}
    \acro{JSON}{JSON}
    \acro{HCI}{Human-Computer Interaction}
    \acro{ARIA}{Accessible Rich Internet Application}
    \acro{W3C}{World Wide Web Consortium}
    \acro{WCAG}{Web Content Accessibility Guidelines}
    \acro{WAI}{Web Accessibility Initiative}
    \acro{JAWS}{Job Access With Speech}
    \acro{NVDA}{NonVisual Desktop Access}
    \acro{CHI}{Conference on Human Factors in Computing Systems}
    \acro{BLV}{Blind Low-Vision}
    \acro{AT}{Assistive Technology}
    \acro{BANA}{Braille Authority of North America}
    \acro{APH}{American Printing House for the Blind}
    \acro{CS}{Computer Science}
    \acro{UD}{Universal Design}
    \acro{PD}{Participatory Design}
    \acro{ID}{Inclusive Design}
    \acro{PWD}{People with Disabilities}
    \acro{CV}{Computer Vision}
    \acro{NLP}{Natural Language Processing}
    \acro{NVDA}{NonVisual Desktop Access}
    \acro{STEM}{science, technology, engineering, and math}
    \acro{COVID19}{COVID-19}
    \acro{NLIs}{Natural Language Interfaces}
    \acro{SVT}{Sentence Verification Technique}
\end{acronym}

\providecommand{\sectionlabel}[1]{(\S~\ref{#1})}
\ieeedoi{10.1109/TVCG.2021.3114770}

\title{
Accessible Visualization via Natural Language Descriptions: \\A Four-Level Model of Semantic Content
}

\author{
Alan Lundgard and Arvind Satyanarayan}
\authorfooter{
\item Alan Lundgard is with MIT CSAIL. E-mail: lundgard@mit.edu.
\item Arvind Satyanarayan is with MIT CSAIL. E-mail: arvindsatya@mit.edu.
}

\shortauthortitle{Lundgard and Satyanarayan: Accessible Visualization via Natural Language Descriptions}

\abstract{
Natural language descriptions sometimes accompany visualizations to better communicate and contextualize their insights, and to improve their accessibility for readers with disabilities. However, it is difficult to evaluate the usefulness of these descriptions, and how effectively they improve access to meaningful information, because we have little understanding of the semantic content they convey, and how different readers receive this content. In response, we introduce a conceptual model for the semantic content conveyed by natural language descriptions of visualizations. Developed through a grounded theory analysis of 2,147 sentences, our model spans four levels of semantic content: enumerating visualization construction properties (e.g., marks and encodings); reporting statistical concepts and relations (e.g., extrema and correlations); identifying perceptual and cognitive phenomena (e.g., complex trends and patterns); and elucidating domain-specific insights (e.g., social and political context). To demonstrate how our model can be applied to evaluate the effectiveness of visualization descriptions, we conduct a mixed-methods evaluation with 30 blind and 90 sighted readers, and find that these reader groups differ significantly on which semantic content they rank as most useful. Together, our model and findings suggest that access to meaningful information is strongly reader-specific, and that research in automatic visualization captioning should orient toward descriptions that more richly communicate overall trends and statistics, sensitive to reader preferences. Our work further opens a space of research on natural language as a data interface coequal with visualization.
} % end of abstract


\vgtcinsertpkg

\teaser{
    \centering
    \includegraphics[width=\linewidth]{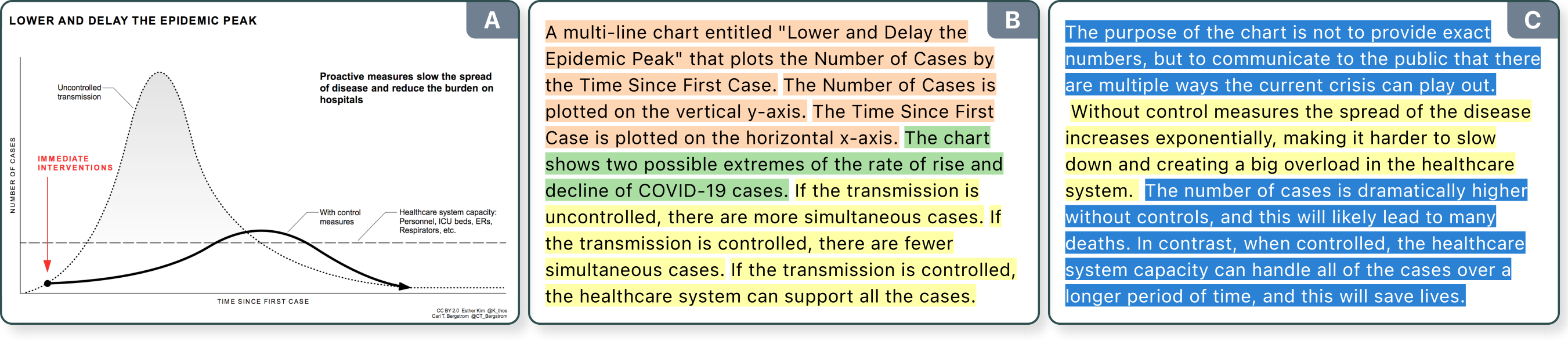}
    \vspace{-8mm}
    \caption{
        Visualizations like ``Flatten the Curve'' (A) efficiently communicate critical public health information, while simultaneously excluding people with disabilities~\cite{bergstrom_sars-cov-2_2020, ehrenkranz_vital_2020}. To promote accessible visualization via natural language descriptions (B, C), we introduce a four-level model of semantic content. Our model categorizes and color codes sentences according to the semantic content they convey.
    }
    \label{fig:teaser}
}

\begin{document}

%% The ``\maketitle'' command must be the first command after the
%% ``\begin{document}'' command. It prepares and prints the title block.

%% the only exception to this rule is the \firstsection command
% \firstsection{Introduction}
\maketitle

\section{Introduction} %for journal use above \firstsection{..} instead
% \section{Introduction}
\label{section:1}

The proliferation of visualizations during the {\smaller COVID-19} pandemic has underscored their double-edged potential: efficiently communicating critical public health information\,---\,as with the immediately-canonical ``Flatten the Curve'' chart (Fig.~\ref{fig:teaser})\,---\,while simultaneously excluding people with disabilities.
\say{\emph{For many people with various types of disabilities, graphics and the information conveyed in them is hard to read and understand,}} says software engineer Tyler Littlefield~\cite{ehrenkranz_vital_2020}, who built a popular text-based {\smaller COVID-19} statistics tracker after being deluged with inaccessible infographics~\cite{sutton_accessible_2020, littlefield_covid-19_2020}.
While natural language descriptions sometimes accompany visualizations in the form of chart captions or alt text (short for \say{alternative text}), these practices remain rare.
Technology educator and researcher Chancey Fleet notes that infographics and charts usually lack meaningful and detailed descriptions, leaving disabled people with \say{\emph{a feeling of uncertainty}} about the pandemic~\cite{ehrenkranz_vital_2020}.
For readers with visual disabilities (approximately 8.1 million in the United States and 253 million worldwide~\cite{ackland_world_2017}), inaccessible visualizations are, at best, demeaning and, at worst, damaging to health, if not accompanied by meaningful and up-to-date alternatives.

Predating the pandemic, publishers and education specialists have long suggested best practices for accessible visual media, including guidelines for tactile graphics~\cite{hasty_guidelines_2011} and for describing ``complex images'' in natural language~\cite{w3c_wai_2019, gould_effective_2008}.
While valuable, visualization authors have yet to broadly adopt these practices, for lack of experience with accessible media, if not a lack of attention and resources.
Contemporary visualization research has primarily attended to color vision deficiency~\cite{chaparro_applications_2017,nunez_optimizing_2018, oliveira_towards_2013}, and has only recently begun to engage with non-visual alternatives~\cite{choi_visualizing_2019, lundgard_sociotechnical_2019} and with accessibility broadly~\cite{kim_accessible_2021, wu_understanding_2021}.
Parallel to these efforts, computer science researchers have been grappling with the engineering problem of automatically generating chart captions~\cite{demir_summarizing_2012, obeid_chart--text_2020, qian_formative_2020}.
While well-intentioned, these methods usually neither consult existing accessibility guidelines, nor do they evaluate their results empirically with their intended readership.
As a result, it is difficult to know how useful (or not) the resultant captions are, or how effectively they improve access to meaningful information.

In this paper, we make a two-fold contribution.
First, we extend existing accessibility guidelines by introducing a conceptual model for categorizing and comparing the semantic content conveyed by natural language descriptions of visualizations.
Developed through a grounded theory analysis of {2,147} natural language sentences, authored by over 120 participants in an online study~\sectionlabel{section:3}, our model spans four levels of semantic content: enumerating visualization construction properties (e.g., marks and encodings); reporting statistical concepts and relations (e.g., extrema and correlations); identifying perceptual and cognitive phenomena (e.g., complex trends and patterns); and elucidating domain-specific insights (e.g., social and political context)~\sectionlabel{section:4}.
% (Moreover, we offer \emph{computational considerations} for generating natural language sentences at each level, such as access to the visualization specification, backing dataset, human perception, and/or domain-specific knowledge.)
Second, we demonstrate how this model can be applied to evaluate the effectiveness of visualization descriptions, by comparing different semantic content levels and reader groups.
We conduct a mixed-methods evaluation in which a group of {30} blind and {90} sighted readers rank the usefulness of descriptions authored at varying content levels~\sectionlabel{section:5}.
Analyzing the resultant {3,600} ranked descriptions, we find significant differences in the content favored by these reader groups: while both groups generally prefer mid-level semantic content, they sharply diverge in their rankings of both the lowest and highest levels of our model.

These findings, contextualized by readers' open-ended feedback, suggest that access to meaningful information is strongly {reader-specific}, and that captions for blind readers should aim to convey a chart's trends and statistics, rather than solely detailing its low-level design elements or high-level insights.
Our model of semantic content is not only \emph{descriptive} (categorizing what \emph{is} conveyed by visualizations) and \emph{evaluative} (helping us to study what \emph{should} be conveyed to whom) but also \emph{generative}~\cite{beaudouin-lafon_instrumental_2000, beaudouin-lafon_designing_2004}, pointing toward novel multimodal and accessible data representations~\sectionlabel{section:practicalImplications}.
Our work further opens a space of research on natural language as a data interface coequal with the language of graphics~\cite{bertin_semiology_1983}, calling back to the original linguistic and semiotic motivations at the heart of visualization theory and design~\sectionlabel{section:theoreticalImplications}.
\section{Related Work}
\label{section:2}

\acresetall % Reset acronyms so they are printed in full in each new section.

Multiple visualization-adjacent literatures have studied methods for describing charts and graphics through natural language\,---\,including accessible media research, \ac{HCI}, \ac{CV}, and \ac{NLP}. But, these various efforts have been largely siloed from one another, adopting divergent methods and terminologies (e.g., the terms ``caption'' and ``description'' are used inconsistently).
% \footnote{
    % The terms ``caption'' and ``description'' are used inconsistently across these literatures.
    % ``caption'' is often used by researchers working with computer vision methods~\cite{qian_formative_2020, xu_show_2016}, whereas ``description'' is more common in accessible media research and \ac{HCI}~\cite{gould_effective_2008, morash_guiding_2015}.
% }
%     %as well consumer-grade visualization software, such as Tableau
%     % In the latter literatures, ``description'' is preferred to ``caption'' because a visualization description may \emph{contain} a caption.
%     % For example, a visualization may come paired with a natural language caption.
%     % But, when providing a natural language \emph{alternative} to that visualization (specifically, for people with visual disabilities), the caption
%     % % paired with the visualization
%     % may itself be \emph{included in} the description. Thus, ``description'' is conceptualized more broadly than ``caption''.
%     In this paper we use ``captioning'' to refer to automatic methods, and ``describing'' to refer to the more general practice of conveying visualized content through natural language.
%     % Additionally, following Bertin, we use ``content'' to refer to the {concepts and information} that can be conveyed through either natural language, visualization, or both.
% }
Here, we survey the diverse terrain of literatures intersecting visualization and natural language.

%----------------------------------------------------------------
\subsection{Automatic Methods for Visualization Captioning}
%----------------------------------------------------------------

Automatic methods for generating visualization captions broadly fall into two categories: those using \ac{CV} and \ac{NLP} methods when the chart is a rasterized image (e.g., \textsc{jpeg}s or \textsc{png}s); and those using structured specifications of the chart's construction (e.g., grammars of graphics).

\subsubsection{Computer Vision and Natural Language Processing}
Analogous to the long-standing \ac{CV} and \ac{NLP} problem of automatically captioning photographic images~\cite{lin_microsoft_2014, krishna_visual_2017, karpathy_deep_2017}, recent work on visualization captioning has aimed to automatically generate accurate and descriptive natural language sentences for charts~\cite{chen_neural_2019, chen_figure_2019, chen_figure_2020, balaji_chart-text_2018, obeid_chart--text_2020, lai_automatic_2020, qian_generating_2021}.
Following the encoder-decoder framework of statistical machine translation~\cite{vinyals_show_2015, xu_show_2016}, these approaches usually take rasterized images of visualizations as input to a \ac{CV} model (the encoder), which learns the visually salient features for outputting a relevant caption via a language model (the decoder).
Training data consists of \textlangle chart, caption\textrangle\, pairs, collected via web-scraping and crowdsourcing~\cite{qian_formative_2020}, or created synthetically from pre-defined sentence templates~\cite{kahou_figureqa_2018}.
While these approaches are well-intentioned, in aiming to address the engineering problem of \emph{how} to automatically generate natural language captions for charts, they have largely sidestepped the complementary (and prior) question: \emph{which} semantic content should be generated to begin with?
Some captions may be more or less descriptive than others, and different readers may receive different semantic content as more or less useful, depending on their levels of data literacy, domain-expertise, and/or visual perceptual ability~\cite{macleod_understanding_2017, morash_guiding_2015, moraes_evaluating_2014}.
To help orient work on automatic visualization captioning, our four-level model of semantic content offers a means of asking and answering these more human-centric questions.

\subsubsection{Structured Visualization Specifications}
In contrast to rasterized images of visualizations, chart templates~\cite{team_bokeh_2014}, component-based architectures~\cite{geveci_vtk_2012}, and grammars of graphics~\cite{satyanarayan_vega-lite_2017} provide not only a structured representation of the visualization's construction, but typically render the visualization in a structured manner as well.
For instance, most of these approaches either render the output visualization as \ac{SVG} or provide a scenegraph API.
Unfortunately, these output representations lose many of the semantics of the structured input (e.g., which elements correspond to axes and legends, or how nesting corresponds to visual perception).
As a result, most present-day visualizations are inaccessible to people who navigate the web using screen readers.
For example, using Apple's VoiceOver to read D3 charts rendered as \ac{SVG} usually outputs an inscrutable mess of screen coordinates and shape rendering properties.
Visualization toolkits can ameliorate this by leveraging their structured input to automatically add \ac{ARIA} attributes to appropriate output elements, in compliance with the \ac{W3C}'s \ac{WAI} guidelines~\cite{w3c_wai_2019}.
Moreover, this structured input representation can also simplify automatically generating natural language captions through template-based mechanisms, as we discuss in \S~\ref{sec:level1}.

%----------------------------------------------------------------
\subsection{Accessible Media and Human-Computer Interaction}
%----------------------------------------------------------------

While automatic methods researchers often note accessibility as a worthy motivation~\cite{demir_summarizing_2012, obeid_chart--text_2020, qian_formative_2020, qian_generating_2021, elzer_exploring_2005, elzer_browser_2007}, evidently few have collaborated directly with disabled people~\cite{choi_visualizing_2019, moraes_evaluating_2014} or consulted existing accessibility guidelines~\cite{lundgard_sociotechnical_2019}.
Doing so is more common to \ac{HCI} and accessible media literatures~\cite{morris_rich_2018, sharif_understanding_2021}, which broadly separate into two categories corresponding to the relative expertise of the description authors: those authored by {experts} (e.g., publishers of accessible media) and those authored by {non-experts} (e.g., via crowdsourcing or online platforms).

\subsubsection{Descriptions Authored by Experts}
Publishers have developed guidelines for describing graphics appearing in \ac{STEM} materials~\cite{gould_effective_2008, benetech_making_nodate}.
Developed by and for authors with some expert accessibility knowledge, these guidelines provide best practices for conveying visualized content in traditional media (e.g., printed textbooks, audio books, and tactile graphics).
But, many of their prescriptions\,---\,particularly those relating to the \emph{content} conveyed by a chart, rather than the \emph{modality} through which the chart is rendered\,---\,are also applicable to web-based visualizations.
Additionally, web accessibility guidelines from \ac{W3C} provide best-practices for writing descriptions of ``complex images'' (including canonical chart types), either in a short description alt text attribute, or as a long textual description displayed alongside the visual image~\cite{w3c_wai_2019}.
While some of these guidelines have been adopted by visualization practitioners~\cite{cesal_writing_2020, schepers_why_2020, elavsky_chartability_2021, fisher_creating_2019, watson_accessible_2018, watson_accessible_2017-1, fossheim_introduction_2020}, we here bring special attention to the empirically-grounded and well-documented guidelines created by the \textsc{wgbh} National Center for Accessible Media~\cite{gould_effective_2008} and by the Benetech Diagram Center~\cite{benetech_making_nodate}.

\subsubsection{Descriptions Authored by Non-Experts}
Frequently employed in \ac{HCI} and visualization research, crowdsourcing is a technique whereby remote non-experts complete tasks currently infeasible for automatic methods, with applications to online accessibility~\cite{bigham_vizwiz_2010}, as well as remote description services like \emph{Be My Eyes}.
For example, Morash et al. explored the efficacy of two types of non-expert tasks for authoring descriptions of visualizations: non-experts authoring free-form descriptions without expert guidance, versus those filling-in sentence templates pre-authored by experts~\cite{morash_guiding_2015}.
While these approaches can yield more richly detailed and ``natural''-sounding descriptions (as we discuss in \S~\ref{section:5}), and also provide training data for auto-generated captions and annotations~\cite{qian_formative_2020, kong_extracting_2014}, it is important to be attentive to potential biases in human-authored descriptions~\cite{bennett_its_2021}.

%----------------------------------------------------------------
\subsection{Natural Language Hierarchies and Interfaces}
%----------------------------------------------------------------

Apart from the above methods for generating descriptions, prior work has adopted linguistics-inspired framings to elucidate how natural language is used to {describe}\,---\,as well as {interact with}\,---\,visualizations.

\subsubsection{Using Natural Language to Describe Visualizations}

Demir et al. have proposed a hierarchy of six syntactic complexity levels corresponding to a set of propositions that might be conveyed by bar charts~\cite{demir_summarizing_2012}.
Our model differs in that it orders \emph{semantic} content\,---\,i.e., \emph{what} meaning the natural language sentence conveys\,---\,rather than \emph{how} it does so syntactically. 
Thus, our model is agnostic to a sentence’s length, whether it contains multiple clauses or conjunctions, which has also been a focus of prior work in automatic captioning~\cite{qian_formative_2020}.
Moreover, whereas Demir et al. speculatively ``envision'' their set of propositions to construct their hierarchy, we arrive to our model empirically through a multi-stage grounded theory process~\sectionlabel{section:3}.
Perhaps closest to our contribution are a pair of papers by Kosslyn~\cite{kosslyn_understanding_1989} and Livingston \& Brock~\cite{livingston_position_2020}.
Kosslyn draws on canonical linguistic theory, to introduce three levels for analyzing charts: the \emph{syntactic} relationship between a visualization elements; the \emph{semantic} meaning of these elements in what they depict or convey; and the \emph{pragmatic} aspects of what these elements convey in the broader context of their reading~\cite{kosslyn_understanding_1989}.
We seeded our model construction with a similar linguistics-inspired framing, but also evaluated it empirically, to further decompose the semantic levels~\sectionlabel{section:3.1}.
Livingston \& Brock adapt Kosslyn's ideas to generate what they call ``visual sentences'': natural language sentences that are the result of executing a single, specific analytic task against a visualization~\cite{livingston_position_2020}.
Inspired by the Sentence Verification Technique (\textsc{svt})~\cite{royer_sentence_1979, royer_developing_2001}, this work considers visual sentences for assessing graph comprehension, hoping to offer a more ``objective'' and automated alternative to existing visualization literacy assessments~\cite{lee_vlat_2016, galesic_graph_2011}.
While we adopt a more qualitative process for constructing our model, Livingston \& Brock's approach suggests opportunities for future work: might our model map to similarly-hierarchical models of analytic tasks~\cite{brehmer_multi-level_2013, amar_low-level_2005}?

\subsubsection{Using Natural Language to Interact with Visualizations}

Adjacently, there is a breadth of work on \ac{NLIs} for constructing and exploring visualizations~\cite{narechania_nl4dv_2021, hearst_toward_2019, setlur_inferencing_2019, kim_answering_2020}.
While our model primarily considers the natural language sentences that are \emph{conveyed by} visualizations (cf., natural language as \emph{input} for chart specification and exploration)~\cite{srinivasan_collecting_2021}, our work may yet have implications for \ac{NLIs}.
For example, Hearst et al. have found that many users of chatbots prefer \emph{not} to see charts and graphics alongside text in the conversational dialogue interface~\cite{hearst_would_2019}.
By helping to decouple visual-versus-linguistic data representations, our model might be applied to offer these users a textual alternative to inline charts.
Thus, we view our work as complementary to \ac{NLIs}, facilitating multimodal and more accessible data representations~\cite{kim_facilitating_2018}, while helping to clarify the theoretical relationship between charts and captions~\cite{kim_towards_2021, ottley_curious_2019}, and other accompanying text~\cite{xiong_curse_2020, kong_frames_2018, kong_trust_2019, adar_communicative_2020}.
\section{Constructing the Model: Employing the Grounded Theory Methodology}
\label{section:3}

To construct our model of semantic content we conducted a multi-stage process, following the \emph{grounded theory} methodology.
Often employed in \ac{HCI} and the social sciences, grounded theory offers a rigorous method for making sense of a domain that lacks a dominant theory, and for constructing a new theory that accounts for diverse phenomena within that domain~\cite{olson_curiosity_2014}.
The methodology approaches theory construction \emph{inductively}\,---\,through multiple stages of inquiry, data collection, ``coding'' (i.e., labeling and categorizing), and refinement\,---\,as well as \emph{empirically}, remaining strongly based (i.e., ``grounded'') in the data~\cite{olson_curiosity_2014}.
To construct our model of semantic content, we proceeded in two stages.
First, we conducted small-scale data collection and initial open coding to establish preliminary categories of semantic content.
Second, we gathered a larger-scale corpus to iteratively refine those categories, and to verify their coverage over the space of natural language descriptions.

\subsection{Initial Open Coding}
\label{section:3.1}

We began gathering preliminary data by searching for descriptions accompanying visualizations in journalistic publications (including the websites of \emph{FiveThirtyEight}, the \emph{New York Times} and the \emph{Financial Times}), but found that these professional sites usually provided no textual descriptions\,---\,neither as a caption alongside the chart, nor as alt text for screen readers.
Indeed, often these sites were engineered so that screen readers would pass over the visualizations entirely, as if they did not appear on the page at all.
Thus, to proceed with the grounded theory method, we conducted initial \emph{open coding} (i.e., making initial, qualitative observations about our data, in an  ``open-minded'' fashion) by studying preliminary data from two sources.
We collected 330 natural language descriptions from over 100 students enrolled in a graduate-level data visualization class.
As a survey-design pilot to inform future rounds of data collection~\sectionlabel{section:study1design}, these initial descriptions were collected with minimal prompting: students were instructed to simply ``describe the visualization'' without specifying what kinds of semantic content that might include.
The described visualizations covered a variety of chart types (e.g., bar charts, line charts, scatter plots) as well as dataset domains (e.g., public health, climate change, and gender equality).
To complement the student-authored descriptions, from this same set of visualizations, we curated a set of 20 and wrote our (the authors') own descriptions, attempting to be as richly descriptive as possible.
Throughout, we adhered to a linguistics-inspired framing by attending to the semantic and pragmatic aspects of our writing: which content could be conveyed through the graphical sign-system alone, and which required drawing upon our individual background knowledge, experiences, and contexts.

Analyzing these preliminary data, we proceeded to the next stage in the grounded theory method: forming \emph{axial codes} (i.e., open codes organized into broader abstractions, with more generalized meaning~\cite{olson_curiosity_2014}) corresponding to different content. We began to distinguish between content about a visualization’s construction (e.g., its title, encodings, legends), content about trends appearing in the visualized data (e.g., correlations, clusters, extrema), and content relevant to the visualized data but not represented in the visualization itself (e.g., explanations based on current events and domain-specific knowledge).
From these axial codes, different \emph{categories} (i.e., groupings delineated by shared characteristics of the content) began to emerge~\cite{olson_curiosity_2014}, corresponding to a chart's encoded elements, latent statistical relations, perceptual trends, and context.
We refined these content categories iteratively by first writing down descriptions of new visualizations (again, as richly as possible), and then attempting to categorize each sentence appearing in that description.
If we encountered a sentence that didn't fit within any category, we either refined the specific characteristics belonging to an existing category, or we created a new category, where appropriate.

%%%%%%%%%%%%%%%%%%%%%%%%%%%%%%%%%%%%%%%%%%%%%%%%%%%%%%%%%%%%%%%%%
\begin{table}
\caption{Breakdown of the 50 curated visualizations, across the three dimensions: type, topic, and difficulty. (N.b., each column sums to 50.)}
\centering

\begin{tabular}{llllll}
\hline
\textsc{chart type} & & \textsc{topic} & & \textsc{difficulty}  & \\
\hline
\texttt{bar} & 18 & \texttt{academic} & 15 & \texttt{easy} & 21\\
\texttt{line} & 21 & \texttt{business} & 18 & \texttt{medium} & 20\\
\texttt{scatter} & 11 & \texttt{journalism} & 17 & \texttt{hard} & 9\\
\hline
\end{tabular}
\label{table:curatedset}
\vspace{-5mm}
\end{table}
%%%%%%%%%%%%%%%%%%%%%%%%%%%%%%%%%%%%%%%%%%%%%%%%%%%%%%%%%%%%%%%%%

\subsection{Gathering A Corpus}

The prior inductive and empirical process resulted in a set of preliminary content categories.
To test their robustness, and to further refine them, we conducted an online survey to gather a larger-scale corpus of {582} visualization descriptions comprised of {2,147} sentences.

\subsubsection{Survey Design}
\label{section:study1design}

We first curated a set of 50 visualizations drawn from the MassVis dataset~\cite{borkin_what_2013, borkin_beyond_2016}, Quartz’s Atlas visualization platform~\cite{poco_reverse-engineering_2017}, examples from the Vega-Lite gallery~\cite{satyanarayan_vega-lite_2017}, and the aforementioned journalistic publications.
We organized these visualizations along three dimensions: the visualization \emph{type} (bar charts, line charts, and scatter plots); the \emph{topic} of the dataset domain (academic studies, business-related, or non-business data journalism); and their \emph{difficulty} based on an assessment of their visual and conceptual complexity.
We labeled visualizations as ``easy'' if they were basic instances of their canonical type (e.g., single-line or un-grouped bar charts), as ''medium'' if they were more moderate variations on canon (e.g., contained bar groupings, overlapping scatterplot clusters, visual embellishments, or simple transforms), and as ''hard'' if they strongly diverged from canon (e.g., contained chartjunk or complex transforms such as log scales).
To ensure robustness, two authors labeled the visualizations independently, and then resolved any disagreement through discussion.
Table~\ref{table:curatedset} summarizes the breakdown of the 50 visualizations across these three dimensions.

In the survey interface, participants were shown a single, randomly-selected visualization at a time, and prompted to describe it in complete English sentences.
In our preliminary data collection~\sectionlabel{section:3.1}, we found that without explicit prompting participants were likely to provide only brief and minimally informative descriptions (e.g., sometimes simply repeating the chart title and axis labels).
Thus, to mitigate against this outcome, and to elicit richer semantic content, we explicitly instructed participants to author descriptions that did not \emph{only} refer to the chart’s basic elements and encodings (e.g., it’s title, axes, colors) but to also referred to other content, trends, and insights that might be conveyed by the visualization.
To make these instructions intelligible, we provided participants with a few pre-generated sentences enumerating the visualization's basic elements and encodings (e.g., the \begin{smaller}\hlc[Level1]{\textsf{color coded sentences}}\end{smaller} in Table~\ref{table:visualizations} A.1, B.1, C.1), and prompted them to author semantic content \emph{apart from} what was already conveyed by those sentences.
To avoid biasing their responses, participants were \emph{not} told that they would be read by people with visual disabilities.
This prompting ensured that the survey captured a breadth of semantic content, and not only the most readily-apparent aspects of the visualization's construction.

%%%%%%%%%%%%%%%%%%%%%%%%%%%%%%%%%%%%%%%%%%%%%%%%%%%%%%%%%%%%%%%%%
% Visual Fingerprint Vertical
%%%%%%%%%%%%%%%%%%%%%%%%%%%%%%%%%%%%%%%%%%%%%%%%%%%%%%%%%%%%%%%%%
\begin{figure}
    \centering
    \includegraphics[width=\linewidth]{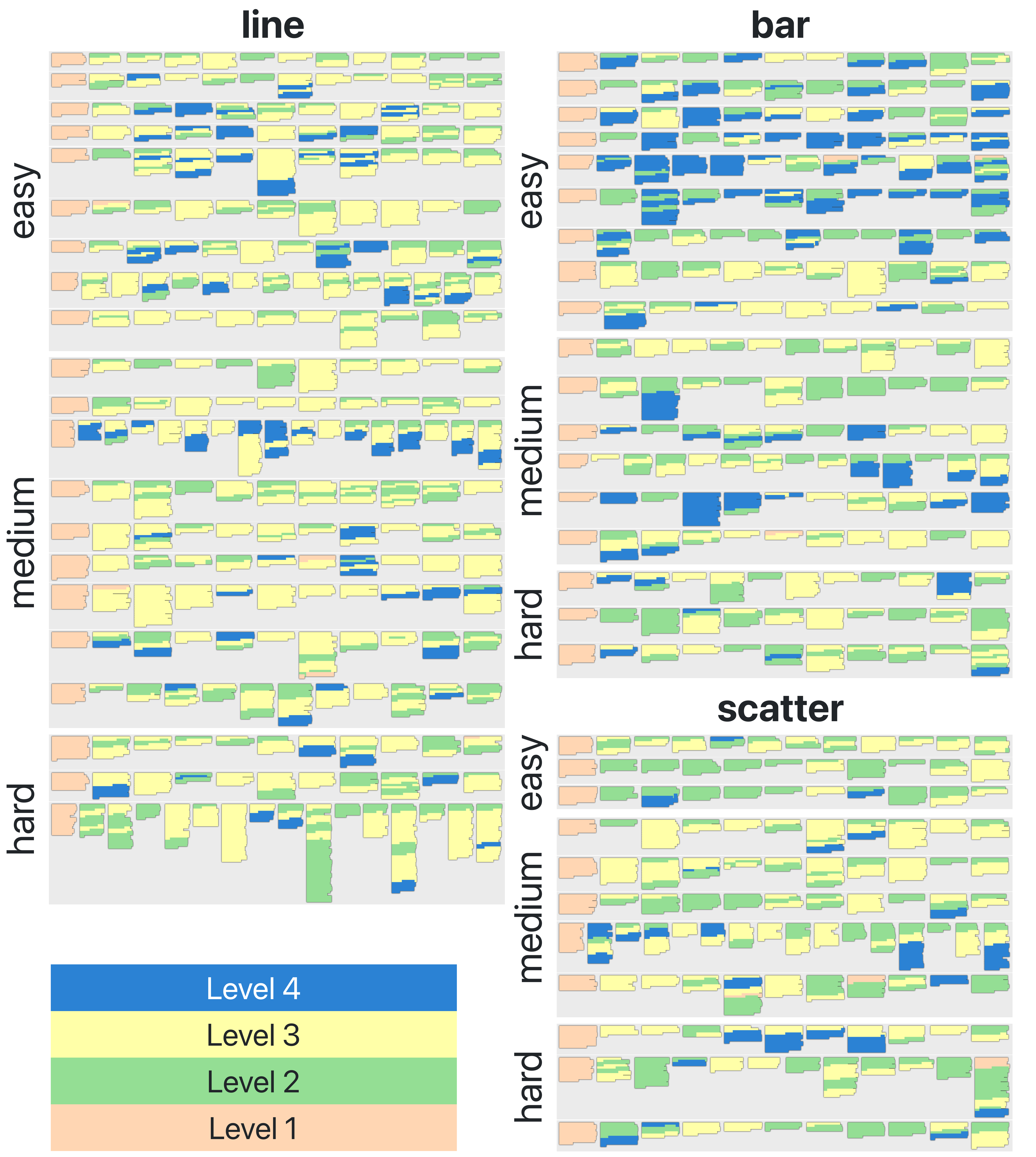}
    \vspace{-4mm}
    \caption{
    A visual ``fingerprint''~\cite{keim_literature_2007} of our corpus, faceted by chart type and difficulty.
    Each row corresponds to a single chart. Each column shows a participant-authored description for that chart, color coded according to our model.
    The first column shows the provided Level 1 prompt.
    }
    \label{fig:fingerprint}
    \vspace{-3mm}
\end{figure}
%%%%%%%%%%%%%%%%%%%%%%%%%%%%%%%%%%%%%%%%%%%%%%%%%%%%%%%%%%%%%%%%%

\subsubsection{Survey Results}
\label{section:study1results}

We recruited 120 survey participants through the \emph{Prolific} platform. In an approximately 30-minute study compensated at a rate of \$10-12 per hour, we asked each participant to describe 5 visualizations (randomly selected from the set of 50), resulting in at least 10 participant-authored descriptions per visualization.
For some visualizations, we collected between 10-15 responses, due to limitations of the survey logic for randomly selecting a visualization to show participants.
% We piloted a shorter survey (asking participants to describe 3 visualizations instead of 6), hypothesizing that this might encourage participants to provide richer, more descriptive content. But, we found that the opposite was true: the length and quality of descriptions decreased appreciably in the 3-visualization study, as compared to the longer 6-visualization study.
% We speculate this may be because the shorter study attracted participants primarily interested in completing quick ``click-through'' tasks.
% Thus, the longer length of our 6-visualization survey may have acted as a ``filter'' to select for the participants willing to more-carefully author descriptions.
In total, this survey resulted in {582} individual descriptions comprised of {2,147} natural language sentences.
We manually cleaned each sentence to correct errors in spelling, grammar, punctuation (n.b., we did not alter the semantic content conveyed by each sentence).
We then labeled each sentence according to the content \emph{categories} developed through our prior grounded theory process.
As before, to ensure robustness, two authors labeled each sentence independently, and then resolved any disagreement through discussion.
This deliberative and iterative process helped us to further distinguish and refine our categories.
For example, we were able to more precisely draw comparisons between sentences reporting computable ``data facts''~\cite{srinivasan_augmenting_2019, wang_datashot_2020} through rigid or templatized articulation (such as \say{\emph{\texttt{[x-encoding]} is positively correlated with \texttt{[y-encoding]}}}), with sentences conveying the same semantic content through more ``natural''-sounding articulation (such as \say{\emph{for the most part, as \texttt{[x-encoding]} increases, so too does \texttt{[y-encoding]}}}).

In summary, the entire grounded theory process resulted in four distinct semantic content categories, which we organize into \emph{levels} in the next section.
A visual ``fingerprint''~\cite{keim_literature_2007} shows how semantic content is distributed across sentences in the corpus (Fig.~\ref{fig:fingerprint}).
Level 1 (consisting of a chart's basic elements and encodings) represents 9.1\% of the sentences in the corpus.
This is expected, since Level 1 sentences were pre-generated and provided as a prompt to our survey participants, as we previously discussed.
The distribution of sentences across the remaining levels is as follows: Level 2 (35.1\%), Level 3 (42.9\%), and Level 4 (12.9\%).
The fairly-balanced distribution suggests that our survey prompting successfully captured natural language sentences corresponding to a breadth of visualized content.

\section{A Four-Level Model of Semantic Content}
\label{section:4}

%%%%%%%%%%%%%%%%%%%%%%%%%%%%%%%%%%%%%%%%%%%%%%%%%%%%%%%%%%%%%%%%%
% Model Table
%%%%%%%%%%%%%%%%%%%%%%%%%%%%%%%%%%%%%%%%%%%%%%%%%%%%%%%%%%%%%%%%%
% Hierarchy Table
%%%%%%%%%%%%%%%%%%%%%%%%%%%%%%%%%%%%%%%%%%%%%%%%%%%%%%%%%%%%%%%%%
\bgroup
\begin{table*}
\caption{A four-level model of semantic content for accessible visualization.
Levels are defined by the semantic content conveyed by natural language descriptions of visualizations.
Additionally, we offer computational considerations for generating the semantic content at each level of the model.
}

\centering
\def\arraystretch{1.5}

\begin{tabular}{m{0.2cm} m{2.5cm} m{5.5cm} m{6cm}}
\hline
\textsc{\#}
& \textsc{level keywords}
& \textsc{semantic content}
& \textsc{computational considerations} \\
\hline

\cellcolor{Level4}\textcolor{white}{\texttt{4}}
& \cellcolor{Level4Light}\texttt{contextual and domain-specific}
& \cellcolor{Level4Light}{\em domain-specific \mbox{insights}, current events, \mbox{social} and \mbox{political} context, explanations}
& \cellcolor{Level4Light}{contextual knowledge and domain-specific \mbox{expertise} (\emph{perceiver-dependent})}
\\

\cellcolor{Level3}\texttt{3}
& \cellcolor{Level3Light}\texttt{perceptual and cognitive}
& \cellcolor{Level3Light}{\em complex trends, pattern synthesis, \mbox{exceptions}, commonplace concepts}
& \cellcolor{Level3Light}{reference to the rendered visualization and \mbox{``common} knowledge'' (\emph{perceiver-dependent})}
\\

\cellcolor{Level2}\texttt{2}
& \cellcolor{Level2Light}\texttt{statistical and relational}
& \cellcolor{Level2Light}{\em descriptive statistics, extrema, outliers, \mbox{correlations}, point-wise comparisons}
& \cellcolor{Level2Light}{access to the visualization specification or \mbox{backing} dataset (\emph{perceiver-independent})}
\\

\cellcolor{Level1}{\texttt{1}}
& \cellcolor{Level1Light}\texttt{elemental and encoded}
& \cellcolor{Level1Light}{\em chart type, encoding channels, title, axis ranges, labels, colors}
& \cellcolor{Level1Light}{access to the visualization specification or \mbox{rasterized} image (\emph{perceiver-independent})}
\\

\hline
\end{tabular}
\label{table:hierarchy}
\vspace{-5mm}
\end{table*}
\egroup
%%%%%%%%%%%%%%%%%%%%%%%%%%%%%%%%%%%%%%%%%%%%%%%%%%%%%%%%%%%%%%%%%

%%%%%%%%%%%%%%%%%%%%%%%%%%%%%%%%%%%%%%%%%%%%%%%%%%%%%%%%%%%%%%%%%

Our grounded theory process yielded a four-level model of semantic content for the natural language description of visualizations. 
In the following subsections, we introduce the levels of the model and provide example sentences for each.
Table~\ref{table:hierarchy} summarizes the levels, and Table~\ref{table:visualizations} shows example visualizations from our corpus and corresponding descriptions, color coded according to the model's color scale.
Additionally, we offer practical \emph{computational considerations} regarding the feasibility of generating sentences at each level, with reference to the present-day state-of-the-art methods described in Related Work.
While we present them alongside each other for ease of explication, we emphasize that the model levels and computational considerations are theoretically decoupled: the model is indexed to the {semantic content} conveyed by natural language sentences, not to the {computational means} through which those sentences may or may not be generated.

%%%%%%%%%%%%%%%%%%%%%%%%%%%%%%%%%%%%%%%%%%%%%%%%%%%%%%%%%%%%%%%%%
% LEVEL 1
%%%%%%%%%%%%%%%%%%%%%%%%%%%%%%%%%%%%%%%%%%%%%%%%%%%%%%%%%%%%%%%%%
\subsection{Level 1: Elemental and Encoded Properties}
\label{sec:level1}

At the first level, there are sentences whose semantic content refers to elemental and encoded properties of the visualization (i.e., the visual components that comprise a graphical representation's design and construction).
These include the chart type (bar chart, line graph, scatter plot, etc.), its title and legend, its encoding channels, axis labels, and the axis scales. Consider the following sentence (Table~\ref{table:visualizations}.A.1).

\begin{quote}
    \vspace{-1mm}
    \small
    \hlc[Level1]{\textsf{Mortality rate is plotted on the vertical y-axis from 0 to 15\%. Age is plotted on the horizontal x-axis in bins: 10-19, 20-29, 30-39, 40-49, 50-59, 60-69, 70-79, 80+.}}
    \vspace{-1mm}
\end{quote}

\noindent
This sentence ``reads off'' the axis labels and scales as they appear in the bar chart, with no additional synthesizing or interpretation.
Sentences such as this are placed at the lowest level in the model because they refer to content that is \emph{foundational} to visualization construction---comprising the elemental properties of the ``language'' of graphics~\cite{bertin_semiology_1983}.

%----------------------------------------------------------------
% L1 Computation
%----------------------------------------------------------------
\textbf{\emph{Computational Considerations.}}
Semantic content at Level 1 is so foundational that it has long been formalized\,---\,not only theoretically, as in Bertin's \emph{Semiology of Graphics}, but also mathematically and programmatically, as a ``grammar of graphics'' that precisely defines the algorithmic rules for constructing canonical chart types.~\cite{wilkinson_grammar_2005}.
In the case of these construction grammars, Level 1 content is \emph{directly encoded} in the visualization's structured specification (i.e., mappings between data fields and visual properties)~\cite{satyanarayan_vega-lite_2017}.
Thus, for these grammars, generating sentences at Level 1 can amount to ``filling in the blank'' for a pre-defined sentence template. For example, given an appropriate template, the following natural language sentence could be trivially computed using the data encoded in the visualization specification.
\begin{displayquote}
    % \small
    \vspace{-1mm}
    \say{\emph{This is a \texttt{[chart-type]} entitled \texttt{[chart-title]}.
    \texttt{[y-encoding]} is plotted on the vertical y-axis from \texttt{[y-min]} to \texttt{[y-max]}. \texttt{[x-encoding]} is plotted on the horizontal x-axis from \texttt{[x-min]} to \texttt{[x-max]}}.}
    \vspace{-1mm}
\end{displayquote}
\noindent
And similarly, for other sentence templates and elemental properties encoded in the visualization's structured specification.
If the structured specification is not available, however, or if it does not follow a declarative grammar, then \ac{CV} and \ac{NLP} methods have also shown promise when applied to rasterized visualization images (e.g., \textsc{jpeg}s or \textsc{png}s).
For example, recent work has shown that Level 1 semantic content can be feasibly generated provided an appropriate training dataset of pre-defined sentence templates~\cite{kahou_figureqa_2018}, or by extracting a visualization's structured specification from a rasterized visualization image~\cite{poco_reverse-engineering_2017}.

%%%%%%%%%%%%%%%%%%%%%%%%%%%%%%%%%%%%%%%%%%%%%%%%%%%%%%%%%%%%%%%%%
% LEVEL 2
%%%%%%%%%%%%%%%%%%%%%%%%%%%%%%%%%%%%%%%%%%%%%%%%%%%%%%%%%%%%%%%%%
\subsection{Level 2: Statistical Concepts and Relations}

At the second level, there are sentences whose semantic content refers to abstract statistical concepts and relations that are latent the visualization’s backing dataset.
This content conveys computable descriptive statistics (such as mean, standard deviation, extrema, correlations)\,---\,what have sometimes been referred to as ``data facts'' because they are ``objectively'' present within a given dataset~\cite{srinivasan_augmenting_2019, wang_datashot_2020} (as opposed to primarily observed via visualization, which affords more opportunities for subjective interpretation).
In addition to these statistics, Level 2 content includes \emph{relations} between data points (such as ``greater than'' or ``lesser than'' comparisons).
Consider the following sentences (Table~\ref{table:visualizations}.C.2).
\begin{quote}
    \small
    \vspace{-1mm}
    \hlc[Level2]{\textsf{For low income countries, the average life expectancy is 60 years for men and 65 years for women. For high income countries, the average life expectancy is 77 years for men and 82 years for women.}}
    \vspace{-1mm}
\end{quote}

\noindent
These two sentences refer to a statistical property: the computed mean of the life expectancy of a population, faceted by gender and country income-level.
Consider another example (Table~\ref{table:visualizations}.A.2).

\begin{quote}
    \small
    \vspace{-1mm}
     \hlc[Level2]{\textsf{The highest COVID-19 mortality rate is in the 80+ age range, while the lowest mortality rate is in 10-19, 20-29, 30-39, sharing the same rate.}}
    \vspace{-1mm}
\end{quote}

\noindent
Although this sentence is more complex, it nevertheless resides at Level 2. It refers to the \emph{extrema} of the dataset (i.e., the ``highest'' and ``lowest'' mortality rates), and makes two \emph{comparisons} (i.e., a comparison between the extrema, and another between age ranges sharing the lowest mortality rate).
All of the above sentences above share the same characteristic, distinguishing them from those at Level 1: they refer to \emph{relations} between points in the dataset, be they descriptive statistics or point-wise comparisons.
Whereas Level 1 sentences ``read off'' the visualization's elemental properties, Level 2 sentences ``report'' statistical concepts and relations within the chart's backing dataset.

%----------------------------------------------------------------
% L2 Computation
%----------------------------------------------------------------
\textbf{\emph{Computational Considerations.}}
While semantic content at Level 1 requires \emph{only} reference to the visualization's specification, content at Level 2 \emph{also} requires access to the backing dataset.
Here, the two categories of automatic methods begin to diverge in their computational feasibility.
For visualizations with a structured specification, generating sentences at Level 2 is effectively as easy as generating sentences at Level 1: it requires little more computation to calculate and report descriptive statistics when the software has access to the backing dataset (i.e., encoded as part of the visualization specification).
Indeed, many visualization software systems (such as Tableau's Summary Card, Voder~\cite{srinivasan_augmenting_2019}, Quill \textsc{nlg} Plug-In for Power BI, and others) automatically compute summary statistics and present them in natural language captions.
By contrast, for \ac{CV} and \ac{NLP} methods, generating Level 2 sentences from a rasterized image is considerably more difficult\,---\,although not entirely infeasible\,---\,depending on the chart type and complexity.
For example, these methods can sometimes report extrema (e.g., which age ranges exhibit the highest and lowest mortality rates in~\ref{table:visualizations}.A.2)~\cite{obeid_chart--text_2020, demir_generating_2008}.
Nevertheless, precisely reporting descriptive statistics (e.g., the computed mean of points in a scatter plot) is less tractable, without direct access to the chart's backing dataset.

%%%%%%%%%%%%%%%%%%%%%%%%%%%%%%%%%%%%%%%%%%%%%%%%%%%%%%%%%%%%%%%%%
% Visualization & Description Table
%%%%%%%%%%%%%%%%%%%%%%%%%%%%%%%%%%%%%%%%%%%%%%%%%%%%%%%%%%%%%%%%%
% Visualization & Description Table
%%%%%%%%%%%%%%%%%%%%%%%%%%%%%%%%%%%%%%%%%%%%%%%%%%%%%%%%%%%%%%%%%
\bgroup
\begin{table*}
\caption{Example visualizations and descriptions from our corpus. Paragraph breaks in rows \texttt{A} and \texttt{B} indicate a description authored by a unique participant from our corpus gathering survey~\sectionlabel{section:study1design}, while row \texttt{C} shows an curated exemplar description from our evaluation~\sectionlabel{section:study2design}.
% We include the entire corpus, and all of the curated exemplars used in the evaluation, in the supplementary material.
}
\centering
\def\arraystretch{1.5}

\begin{tabular}{m{0.2cm} m{7.5cm} m{9cm}}
\hline
\multicolumn{2}{l}{\textsc{visualization}} & \textsc{description}
\\
\hline

\cellcolor{GrayLight}\textbf{\texttt{A}} &
\includegraphics[scale=0.33]{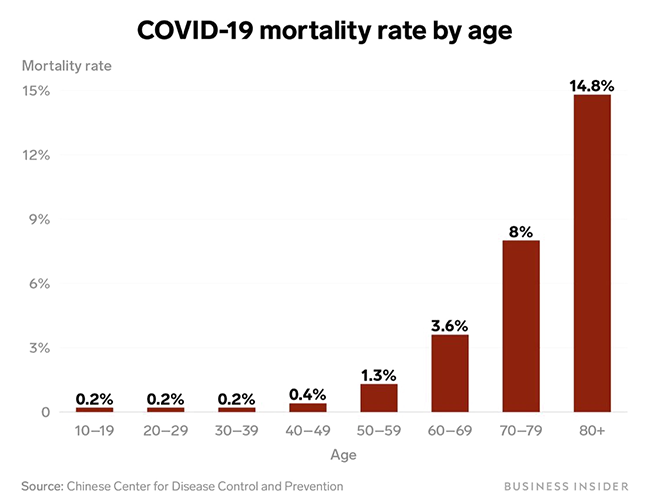}
\textbf{\texttt{[bar, easy, journalism]}}
&
% \cellcolor{GrayLight}
{\sffamily\smaller

    [1]~\hlc[Level1]{This is a vertical bar chart entitled ``COVID-19 mortality rate by age'' that plots Mortality rate by Age. Mortality rate is plotted on the vertical y-axis from 0 to 15\%. Age is plotted on the horizontal x-axis in bins: 10-19, 20-29, 30-39, 40-49, 50-59, 60-69, 70-79, 80+.}
    [2]~\hlc[Level2]{The highest COVID-19 mortality rate is in the 80+ age range, while the lowest mortality rate is in 10-19, 20-29, 30-39, sharing the same rate.}
    [3]~\hlc[Level3]{COVID-19 mortality rate does not linearly correspond to the demographic age.}

    [4]~\hlc[Level3]{The mortality rate increases with age, especially around 40-49 years and upwards.}
    [5]~\textcolor{white}{\hlc[Level4]{This relates to people's decrease in their immunity and the increase of co-morbidity with age.}}

    [6]~\hlc[Level3]{The mortality rate increases exponentially with older people.}
    [7]~\hlc[Level2]{There is no difference in the mortality rate in the range between the age of 10 and 39.}
    [8]~\hlc[Level3]{The range of ages between 60 and 80+ are more affected by COVID-19.}

    [9]~\textcolor{white}{\hlc[Level4]{We can observe that the mortality rate is higher starting at 50 years old due to many complications prior.}}
    [10]~\hlc[Level3]{As we decrease the age, we also decrease the values in mortality by a lot, almost to none.}
}
\\
\hline

\cellcolor{GrayLight}\textbf{\texttt{B}} &
\includegraphics[scale=0.53]{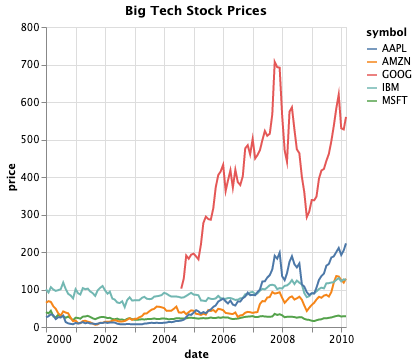}
\textbf{\texttt{[line, medium, business]}}
& 
% \cellcolor{GrayLight}
{\sffamily\smaller

    [1]~\hlc[Level1]{This is a multi-line chart entitled ``Big Tech Stock Prices'' that plots price by date. The corporations include AAPL (Apple), AMZN (Amazon), GOOG (Google), IBM (IBM), and MSFT (Microsoft). The years are plotted on the horizontal x-axis from 2000 to 2010 with an increment of 2 years. The prices are plotted on the vertical y-axis from 0 to 800 with an increment of 200.}
    [2]~\hlc[Level2]{GOOG has the greatest price over time. MSFT has the lowest price over time.}
    [3]~\hlc[Level3]{Prices of particular Big Tech corporations seem to fluctuate but nevertheless increase over time. Years 2008-2009 are exceptions as we can see an extreme drop in prices of all given corporations.}
    [4]~\textcolor{white}{\hlc[Level4]{The big drop in prices was caused by financial crisis of 2007-2008. The crisis culminated with the bankruptcy of Lehman Brothers on September 15, 2008 and an international banking crisis.}}
    
    [5]~\textcolor{white}{\hlc[Level4]{At the beginning of 2008, every of this stock price went down, likely due to the financial crisis.}}
    [6]~\hlc[Level3]{Then they have risen again and dropped again, more so than previously.}
    
    [7]~\hlc[Level2]{GOOG has the highest price over the years. MSFT has the lowest price over the years.}
    [8]~\hlc[Level3]{GOOG quickly became the richest one of the Big Tech corporations.}
    [9]~\textcolor{white}{\hlc[Level4]{GOOG had experienced some kind of a crisis in 2009, because their prices drop rapidly, but then rebounded.}}
}
\\
\hline

\cellcolor{GrayLight}\textbf{\texttt{C}} &
\includegraphics[scale=0.82]{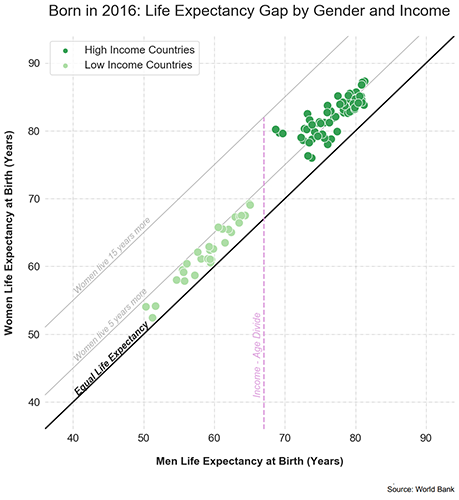}
\textbf{\texttt{[scatter, hard, academic]}}
&
% \cellcolor{GrayLight}
{\sffamily\smaller

    [1]~\hlc[Level1]{This is a scatter plot entitled ``Born in 2016: Life Expectancy Gap by Gender and Income'' that plots Women Life Expectancy at Birth (Years) by Men Life Expectancy at Birth (Years). The Women Life Expectancy at Birth is plotted on the vertical y-axis from 40 to 90 years. The Men Life Expectancy at Birth is plotted on the horizontal x-axis from 40 to 90 years. High Income Countries are plotted in dark green. Low Income Countries are plotted in light green. A 45 degree line from the origin represents Equal Life Expectancy.}
    [2]~\hlc[Level2]{For low income countries, the average life expectancy is 60 years for men and 65 years for women. For high income countries, the average life expectancy is 77 years for men and 82 years for women.}
    [3]~\hlc[Level3]{Overall, women have a slightly higher life expectancy than men. Women live around 5 to 10 years longer than men. The low income countries are more scattered than the high income countries. There is a visible gap between high and low income countries, indicated by the Income-Age Divide line.}
    [4]~\textcolor{white}{\hlc[Level4]{People living in low-income countries tend to have a lower life expectancy than the people living in high-income countries, likely due to many societal factors, including access to healthcare, food, other resources, and overall quality of life. People who live in lower income countries are more likely to experience deprivation and poverty, which can cause related health problems.}}
}
\\
\hline
\end{tabular}
\label{table:visualizations}
\end{table*}
%%%%%%%%%%%%%%%%%%%%%%%%%%%%%%%%%%%%%%%%%%%%%%%%%%%%%%%%%%%%%%%%%

%%%%%%%%%%%%%%%%%%%%%%%%%%%%%%%%%%%%%%%%%%%%%%%%%%%%%%%%%%%%%%%%%

%%%%%%%%%%%%%%%%%%%%%%%%%%%%%%%%%%%%%%%%%%%%%%%%%%%%%%%%%%%%%%%%%
% LEVEL 3
%%%%%%%%%%%%%%%%%%%%%%%%%%%%%%%%%%%%%%%%%%%%%%%%%%%%%%%%%%%%%%%%%
\subsection{Level 3: Perceptual and Cognitive Phenomena}

At the third level, there are sentences whose semantic content refers to perceptual and cognitive phenomena appearing in the visual representation of the data.
When compared to, and defended against, other forms of data analysis (e.g., purely mathematical or statistical methods), visualization is often argued to confer some unique benefit to human readers.
That is, visualizations do not only ``report'' descriptive statistics of the data (as in Level 2), they also \emph{show} their readers something \emph{more}: they surface unforeseen trends, convey complex multi-faceted patterns, and identify noteworthy exceptions that aren't readily apparent through non-visual methods of analysis (cf., Anscombe's Quartet or the Datasaurus Dozen~\cite{matejka_same_2017}).
Level 3 sentences are comprised of content that refers to these perceptual and cognitive phenomena, usually articulated in ``natural''-sounding (rather than templatized) language.
Consider the following examples (Table~\ref{table:visualizations}.B.3 and \ref{table:visualizations}.C.3, respectively).
\begin{quote}
    \small
    \vspace{-1mm}
    \hlc[Level3]{\textsf{Prices of particular Big Tech corporations seem to fluctuate but nevertheless increase over time. Years 2008-2009 are exceptions as we can see an extreme drop in prices of all given corporations.}}
    \vspace{-1mm}
\end{quote}

\begin{quote}
    \small
    \hlc[Level3]{\textsf{The low income countries are more scattered than the high income countries. There is a visible gap between high and low income countries, indicated by the Income-Age Divide line.}}
    \vspace{-1mm}
\end{quote}

\noindent
These sentences convey the ``overall gist'' of complex trends and patterns (e.g., stock prices ``seem to fluctuate but nevertheless increase''), synthesize multiple trends to identify exceptions (e.g., ``years 2008-2009 are exceptions as we can see an extreme drop'' of multiple graphed lines at that point in time), and do so in ``natural''-sounding language, by referencing commonplace concepts (such as ``fluctuate'', ``extreme drop'', ``visible gap'').
N.b., ``natural''-sounding articulation is necessary but insufficient for Level 3 membership, as it is also possible to articulate Level 1 or 2 content in a non-templatized fashion~\sectionlabel{section:study1results}.

%----------------------------------------------------------------
% L3 Computation
%----------------------------------------------------------------
\textbf{\emph{Computational Considerations.}}
At Level 3, we begin to reach and exceed the limits of present-day state-of-the-art automatic methods.
While there exist ``off-the-shelf'' statistical packages for computing basic trends and predictions in a dataset (e.g., correlations, polynomial regressions, statistical inferences), visualizations allow us to perceive and articulate complex trends for which there may exist no line of ``best fit''.
While automatic methods may eventually approach (or exceed) human capabilities on well-defined tasks~\cite{obeid_chart--text_2020}, for now Level 3 semantic content is likely generated via human (rather than machine) perception and cognition~\cite{morash_guiding_2015}.
Taking inspiration from the \say{mind-independent} versus \say{mind-dependent} ontological distinction~\cite{ali_mind-dependent_2016}, we define sentences at Levels 1 and 2 as \emph{perceiver-independent} (i.e., their content can be generated independently of human or machine perception, without reference to the visualization), while sentences at Level 3 are \emph{perceiver-dependent} (i.e., their content requires a perceiver of some sort; likely a human, although machine perception may increasingly suffice for generating Level 3 content).
Table~\ref{table:hierarchy} summarizes this distinction.

%%%%%%%%%%%%%%%%%%%%%%%%%%%%%%%%%%%%%%%%%%%%%%%%%%%%%%%%%%%%%%%%%
% LEVEL 4
%%%%%%%%%%%%%%%%%%%%%%%%%%%%%%%%%%%%%%%%%%%%%%%%%%%%%%%%%%%%%%%%%
\subsection{Level 4: Contextual and Domain-Specific Insights}

Finally, at the fourth level, there are sentences whose semantic content refers to contextual and domain-specific knowledge and experience.
Consider the following two examples (Table~\ref{table:visualizations}.B.4 and~\ref{table:visualizations}.C.4).
\begin{quote}
    \small
    \vspace{-1mm}
    \textcolor{white}{\hlc[Level4]{\textsf{The big drop in prices was caused by financial crisis of 2007-2008. The crisis culminated with the bankruptcy of Lehman Brothers on September 15, 2008 and an international banking crisis.}}}
    \vspace{-1mm}
\end{quote}

\begin{quote}
    \small
    \textcolor{white}{\hlc[Level4]{\textsf{People living in low-income countries tend to have a lower life expectancy than the people living in high-income countries, likely due to many societal factors, including access to healthcare, food, other resources, and overall quality of life.}}}
    \vspace{-1mm}
\end{quote}

\noindent
These sentences convey social and political explanations for an observed trend that depends on an individual reader's subjective knowledge about particular world events: the 2008 financial crisis and global socio-economic trends, respectively.
This semantic content is characteristic of what is often referred to as ``insight'' in visualization research.
Although lacking a precise and agreed-upon definition~\cite{law_what_2020, tang_extracting_2017, north_toward_2006, chang_defining_2009, law_characterizing_2020}, an insight is often an observation about the data that is complex, deep, qualitative, unexpected, and relevant~\cite{yi_understanding_2008}.
Critically, insights depend on individual perceivers, their subjective knowledge, and domain-expertise.
Level 4 is where the breadth of an individual reader’s knowledge and experience is brought to bear in articulating something ``insightful'' about the visualized data.

%----------------------------------------------------------------
% L4 Computation
%----------------------------------------------------------------
\textbf{\emph{Computational Considerations.}}
{As with Levels 3, we say that Level 4 semantic content is perceiver-dependent, but in a stronger sense}. 
This is because (setting aside consideration of hypothetical future ``artificial general intelligence'') generating Level 4 semantic content is at-present a uniquely human endeavor.
Doing so involves synthesizing background knowledge about the world (such as geographic, cultural, and political relationships between countries), contextual knowledge about current events (e.g., the fact that there was a global recession in 2008), and domain-specific knowledge (e.g., expertise in a particular field of research or scholarship).
However, bespoke systems for narrowly-scoped domains (e.g., those auto-generating stock chart annotations using a corpus of human-authored news articles~\cite{hullman_contextifier_2013}) suggest that some Level 4 content might be feasibly generated sooner rather than later.

Lastly, we briefly note that data-driven predictions can belong to either Level 2, 3, or 4, depending on the semantic content contained therein.
For example: a point-wise prediction at Level 2 (e.g., computing a stock's future expected price using the backing dataset); a prediction about future overall trends at Level 3 (e.g., observing that a steadily increasing stock price will likely continue to rise); a prediction involving contextual or domain-specific knowledge at Level 4 (e.g., the outcome of an election using a variety of poll data, social indicators, and political intuition).

\section{Applying the Model: Evaluating the Effectiveness of Visualization Descriptions}
\label{section:5}

The foregoing conceptual model provides a means of making structured comparisons between different levels of semantic content and reader groups.
To demonstrate how it can be applied to evaluate the effectiveness of visualization descriptions (i.e., whether or not they effectively convey meaningful information, and for whom), we conducted a mixed-methods evaluation in which {30} blind and {90} sighted readers first ranked the usefulness of descriptions authored at varying levels of semantic content, and then completed an open-ended questionnaire.

%%%%%%%%%%%%%%%%%%%%%%%%%%%%%%%%%%%%%%%%%%%%%%%%%%%%%%%%%%%%%%%%%
% Evaluation Design
%%%%%%%%%%%%%%%%%%%%%%%%%%%%%%%%%%%%%%%%%%%%%%%%%%%%%%%%%%%%%%%%%
\subsection{Evaluation Design}
\label{section:study2design}

We selected 15 visualizations for the evaluation, curated to be representative of the categories from our prior survey~\sectionlabel{section:3}.
Specifically, we selected 5 visualizations for each of the three dimensions: \emph{type} (bar, line, scatter), \emph{topic} (academic, business, journalism), and \emph{difficulty} (easy, medium, hard).
For every visualization, participants were asked to rank the usefulness of 4 different descriptions, each corresponding to one level of semantic content, presented unlabeled and in random order.
We piloted this rank-choice interface with 10 sighted readers recruited via \emph{Prolific} and 1 blind reader, a non-academic collaborator proficient with Apple's VoiceOver screen reader. 
Based on this pilot, we rewrote the study instructions to be more intelligible to both groups of readers, added an introductory example task to the evaluation, and improved the screen reader accessibility of our interface (e.g., by reordering nested \textsc{dom} elements to be more intuitively traversed by screen reader).

In addition to curating a representative set of visualizations, we also curated descriptions representative of each level of semantic content.
Participant-authored descriptions from our prior survey often did not contain content from all 4 levels or, if they did, this content was interleaved in a way that was not cleanly-separable for the purpose of a ranking task (Fig.~\ref{fig:fingerprint}).
Thus, for this evaluation, we curated and collated sentences from multiple participant-authored descriptions to create exemplar descriptions, such that each text chunk contained \emph{only} content belonging to a single semantic content level.
Table~\ref{table:visualizations}.C shows one such exemplar description, whereas Table~\ref{table:visualizations}.A and B show the original un-collated descriptions.
For each ranking task, readers were presented with a brief piece of contextualizing text, such as the following.
\begin{displayquote}
    % \small
    \vspace{-1mm}
    \say{\emph{Suppose that you are reading an academic paper about how life expectancy differs for people of different genders from countries with different levels of income. You encounter the following visualization.} [Table~\ref{table:visualizations}.C]
    \emph{Which content do you think would be most useful to include in a textual description of this visualization?}}
    \vspace{-1mm}
\end{displayquote}
Additionally, blind readers were presented with a brief text noting that the hypothetically-encountered visualization was inaccessible via screen reader technology.
In contrast to prior work, which has evaluated chart descriptions in terms of ``efficiency,'' ``informativeness,'' and ``clarity''~\cite{gould_effective_2008, obeid_chart--text_2020}, we intentionally left the definition of ``useful'' open to the reader’s interpretation. 
We hypothesize that ``useful'' descriptions may not be necessarily efficient (i.e., they may require lengthy explanation or background context), and that both informativeness and clarity are constituents of usefulness.
In short, ranking ``usefulness'' affords a holistic evaluation metric.
Participants assigned usefulness rankings to each of the 4 descriptions by selecting corresponding radio buttons, {labeled 1 (least useful) to 4 (most useful)}.
In addition to these 4 descriptions, we included a 5th choice as an ``attention check'': a sentence whose content was entirely irrelevant to the chart to ensure participants were reading each description prior to ranking them.
If a participant did not rank the attention check as least useful, we filtered out their response from our final analysis.
{We include the evaluation interfaces and questions with the Supplemental Material.}

%%%%%%%%%%%%%%%%%%%%%%%%%%%%%%%%%%%%%%%%%%%%%%%%%%%%%%%%%%%%%%%%%
% Participants
%%%%%%%%%%%%%%%%%%%%%%%%%%%%%%%%%%%%%%%%%%%%%%%%%%%%%%%%%%%%%%%%%
\subsection{Participants}

Participants consisted of two reader groups: {90} sighted readers recruited through the \emph{Prolific} platform, and {30} blind readers recruited through our friends in the blind community and through a call for participation sent out via Twitter (n.b., in accessibility research, it is common to compare blind and sighted readers recruited through these means~\cite{bigham_effects_2017}).

\subsubsection{Participant Recruitment}

For sighted readers qualifications for participation included English language proficiency and no color vision deficiency, and blind readers were expected to be proficient with a screen reader, such as \ac{JAWS}, \ac{NVDA}, or Apple's VoiceOver.
Sighted readers were compensated at a rate of \$10-12 per hour, for an approximately 20-minute task. 
Blind readers were compensated at a rate of \$50 per hour, for an approximately 1-hour task. 
This difference in task duration was for two reasons. 
First, participants recruited through \emph{Prolific} are usually not accustomed to completing lengthy tasks\,---\,our prior surveys and pilots suggested that these participants might contribute low-quality responses on ``click-through'' tasks if the task duration exceeded 15--20 minutes\,---\,and thus we asked each participant to rank only 5 of the 15 visualizations at a time.
Second, given the difficulty of recruiting blind readers proficient with screen readers, we asked each blind participant to rank all 15 visualizations, and compensated them at a rate commensurate with their difficult-to-find expertise~\cite{lundgard_sociotechnical_2019}.
In this way, we recruited sufficient numbers of readers to ensure that each of the 15 visualization ranking tasks would be completed by 30 participants from both reader groups.

\subsubsection{Participant Demographics}
Among the {30} blind participants, {53}\% (n={16}) reported their gender as male, {36}\% (n={11}) as female, and {3} participants ``preferred not to say.''
The most common highest level of education attained was a Bachelor's degree ({60}\%, n=18), and most readers were between 20\,--\,40 years old ({66}\%, n=20).
The screen reader technology readers used to complete the study was evenly balanced: VoiceOver (n=10), \ac{JAWS} (n=10), \ac{NVDA} (n=9), and ``other'' (n=1).
Among the {90} sighted participants, {69}\% reported their gender as male (n={62}) and {31}\% as female (n={28}).
The most common highest level of education attained was a high school diploma ({42}\%, n=38) followed by a Bachelor's degree ({40}\%, n=36), and most sighted readers were between 20\,--\,30 years old ({64}\%, n=58).

On a 7-point Likert scale [1=strongly disagree, 7=strongly agree], blind participants reported having ``a good understanding of data visualization concepts'' ($\mu=6.3$, $\sigma=1.03$) as well as ``a good understanding of statistical concepts and terminology'' ($\mu=5.90$, $\sigma=1.01$). Sighted participants reported similar levels of understanding: ($\mu=6.7$, $\sigma=0.73$) and ($\mu=5.67$, $\sigma=1.06$), respectively.
Sighted participants also considered themselves to be ``proficient at reading data visualizations'' ($\mu=5.97$, $\sigma=0.89$) and were able to ``read and understand all of the visualizations presented in this study'' ($\mu=6.44$, $\sigma=0.71$).

% Most readers were between ages 30-39 (n=11), followed by ages 40-49 (n=7), ages 20-29 (n=5), and ages 50-59 (n=1).
% The most common reported occupation was Assistive Technology Specialist or Engineer (n=6), with other occupations including Nonprofit Program Director, Applied Linguist, Artist, and Lawyer.
% The screen reader technology participants used to complete the study was evenly balanced: \ac{JAWS} (n=9), \ac{NVDA} (n=8), VoiceOver (n=8), and "Other" (n=1). 

% Most readers were between ages 30-39 (n=11), followed by ages 40-49 (n=7), ages 20-29 (n=5), and ages 50-59 (n=1).
% Reported occupations included Computer Scientist, Graphic Designer, Taxi Driver, German Philologist, Information Technologist, among many others.

%%%%%%%%%%%%%%%%%%%%%%%%%%%%%%%%%%%%%%%%%%%%%%%%%%%%%%%%%%%%%%%%%
% Evaluation Table
%%%%%%%%%%%%%%%%%%%%%%%%%%%%%%%%%%%%%%%%%%%%%%%%%%%%%%%%%%%%%%%%%
% Evaluation Table Vertical
%%%%%%%%%%%%%%%%%%%%%%%%%%%%%%%%%%%%%%%%%%%%%%%%%%%%%%%%%%%%%%%%%
\bgroup
\begin{table}
\centering
\caption{
(Upper) Rankings [1=least useful, 4=most useful] of semantic content at each level of the model, for blind and sighted readers.
The scale encodes the number of times a given level was assigned a given rank by a reader.
Dotted contour lines delineate Regions with a threshold equal to $\mu + \frac{\sigma}{2}$, each labeled with a capital letter  A\,--\,F.
%in its upper-right-hand corner.
(Lower)
% Pair-wise ranking differences between levels. 
Shaded cells indicate significant ranking differences pair-wise between levels.
}

\def\arraystretch{1.5}
\setlength{\tabcolsep}{2pt}
\smaller

\vspace{-1mm}
%---------------------------------------------------------------
% Heatmaps
%---------------------------------------------------------------
\begin{tabular}{cc}
\hline
\hlc[DirectQuoteA]{\normalsize\textsc{blind readers}} & \hlc[DirectQuoteB]{\normalsize\textsc{sighted readers}} \\
\hline
\includegraphics[width=0.22\textwidth]{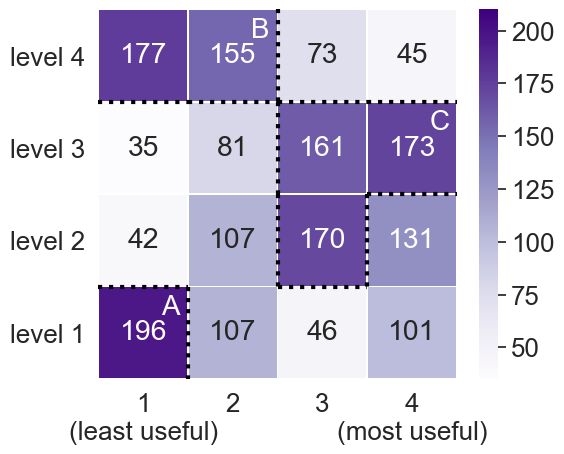}
\vspace{-1mm}
&
\includegraphics[width=0.22\textwidth]{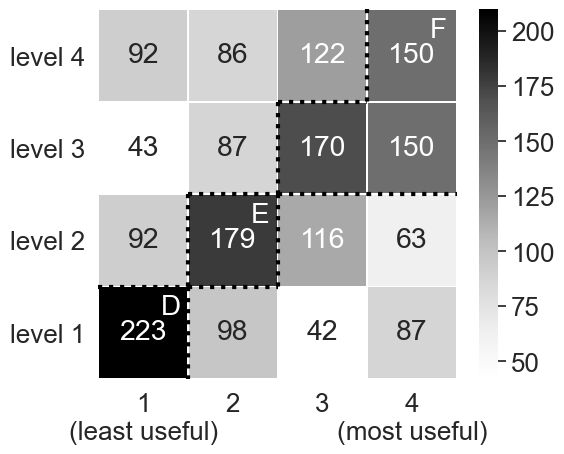}
\\
% \hline
\end{tabular}
% ---------------------------------------------------------------
% ---------------------------------------------------------------
\vspace{6mm}
%---------------------------------------------------------------
% Significance
%---------------------------------------------------------------
\begin{tabular}{c|c|c|c|c|c|c}
\hline
\textsc{levels} & $1 \times 2$ & $1 \times 3$ & $1 \times 4$ & $2 \times 3$ & $2 \times 4$ & $3 \times 4$ \\ \hline
\hlc[DirectQuoteA]{\textsc{blind}} & \cellcolor{yellow!25}$p<0.001$ & \cellcolor{yellow!25}$p<0.001$ & $p<0.321$ & $p<0.148$ & \cellcolor{yellow!25}$p<0.001$ & \cellcolor{yellow!25}$p<0.001$ \\ \hline
\hlc[DirectQuoteB]{\textsc{sighted}} & \cellcolor{yellow!25}$p<0.001$ & \cellcolor{yellow!25}$p<0.001$ & \cellcolor{yellow!25}$p<0.001$ & \cellcolor{yellow!25}$p<0.001$ & \cellcolor{yellow!25}$p<0.001$ & $p<0.059$ \\ \hline
\end{tabular}
%---------------------------------------------------------------
%---------------------------------------------------------------
\label{table:evaluation}
\vspace{-11mm} %FIX ME? Camera-ready
\end{table}
\egroup
\subsection{Quantitative Results}

% \noindent
% \textbf{\emph{Ranked Preferences}}

Quantitative results for the individual rankings ({1,800} per blind and sighted reader groups) are summarized by the heatmaps in Table~\ref{table:evaluation} (Upper), which aggregate the number of times a given content level was assigned a certain rank.
Dotted lines in both blind and sighted heatmaps delineate regions exceeding a threshold\,---\,calculated by taking the mean plus half a standard deviation ($\mu + \frac{\sigma}{2}$) resulting in a value of 139 and 136, respectively\,---\,and are labeled with a capital letter A\,--\,F.

These results exhibit significant differences between reader groups.
For both reader groups, using Friedman’s Test (a non-parametric multi-comparison test for rank-order data) the p-value is $p<0.001$, so we reject the null hypothesis that the mean rank is the same for all four semantic content levels~\cite{garcia_advanced_2010}.
Additionally, in Table~\ref{table:evaluation} (Lower), we find significant ranking differences when making pair-wise comparisons between levels, via Nemenyi’s test (a post-hoc test commonly coupled with Friedman's to make pair-wise comparisons).
There appears to be strong agreement among sighted readers that higher levels of semantic content are more useful: Levels 3 and 4 are found to be most useful (Region~\ref{table:evaluation}.F), while Levels 1 and 2 are least useful (Regions~\ref{table:evaluation}.D and ~\ref{table:evaluation}.E).
Blind readers agree with each other to a lesser extent, but strong trends are nevertheless apparent.
In particular, blind readers rank content and Levels 2 and 3 as most useful (Region~\ref{table:evaluation}.C), and semantic content at Levels 1 and 4 as least useful (Regions~\ref{table:evaluation}.A and ~\ref{table:evaluation}.B).

When faceting these rankings by visualization type, topic, or difficulty we did not observe any significant differences, suggesting that both reader groups rank semantic content levels consistently, regardless of how the chart itself may vary.
Noteworthy for both reader groups, the distribution of rankings for Level 1 is bimodal\,—--\,the only level to exhibit this property.
While a {vast majority} of both blind and sighted readers rank Level 1 content as least useful, this level is ranked ``most useful'' in {101} and {87} instances by blind and sighted readers, respectively. This suggests that both reader groups have a more complicated perspective toward descriptions of a chart's elemental and encoded properties; a finding we explore further by analyzing qualitative data.

%%%%%%%%%%%%%%%%%%%%%%%%%%%%%%%%%%%%%%%%%%%%%%%%%%%%%%%%%%%%%%%%%
% Qualitative Results
%%%%%%%%%%%%%%%%%%%%%%%%%%%%%%%%%%%%%%%%%%%%%%%%%%%%%%%%%%%%%%%%%
\subsection{Qualitative Results}
\label{section:qualResults}

In a questionnaire, we asked readers to use a 7-point Likert scale [1=strongly disagree, 7=strongly agree] to rate their agreement with a set of statements about their experience with visualizations.
We also asked them to offer open-ended feedback about which semantic content they found to be most useful and why.
Here, we summarize the key trends that emerged from these two different forms of feedback, from both \hlc[DirectQuoteA]{blind readers} (BR) and \hlc[DirectQuoteB]{sighted readers} (SR).

%----------------------------------------------------------------
\subsubsection{Descriptions Are Important to Both Reader Groups}
\label{section:qualResultsA}
%----------------------------------------------------------------
% Responding to the statement \say{Providing textual descriptions of data visualizations is important to me}, blind readers expressed a mean level of agreement of $\mu=6.82$ ($SD=0.38$), indicating overwhelming support for visualization descriptions. 
All blind readers reported encountering inaccessible visualizations: either multiple times a week ({43}\%, n=13), everyday ({20}\%, n=6), once or twice a month ({20}\%, n=6), or at most once a week ({17}\%, n=5).
These readers reported primarily encountering these barriers on social media ({30}\%, n=9), on newspaper websites ({13}\%, n=4), and in educational materials ({53}\%, n=16)\,---\,but, most often, barriers were encountered in all of the above contexts ({53}\%, n=16).
% Readers report encountering inaccessible visualizations in each of the following contexts: {30}\% (n=9) on social media (Twitter, Facebook), {13}\% (n=4) on newspaper websites, {3}\% (n=1) in educational materials, and {53}\% (n=16) in all of the above.
Blind readers overwhelmingly agreed with the statements 
\say{I often feel that important public information is inaccessible to me, because it is only available in a visual format} ($\mu=6.1$, $\sigma=1.49$), and \say{Providing textual descriptions of data visualizations is important to me} ($\mu=6.83$, $\sigma=0.38$). 

\begin{displayquote}
% \small
    \vspace{-1mm}
   \hquoteA{I am totally blind, and virtually all data visualizations I encounter are undescribed, and as such are unavailable. This has been acutely made clear on Twitter and in newspapers around the COVID-19 pandemic and the recent U.S. election.
   Often, visualizations are presented with very little introduction or coinciding text. I feel very left out of the world and left out of the ability to confidently traverse that world. The more data I am unable to access, the more vulnerable and devalued I feel.} (BR5)
    \vspace{-1mm}
\end{displayquote}

\noindent
By contrast, sighted readers neither agreed nor disagreed regarding the inaccessibility of information conveyed visually ($\mu=4$, $\sigma=1.57$).
Similarly, they were split on whether they ever experienced barriers to reading visualizations, with {52}\% (n={47}) reporting that they sometimes do (especially when engaging with a \emph{new} topic) and {48}\% (n={43}) reporting that they usually do not.
Nevertheless, sighted readers expressed support for natural language descriptions of visualizations ($\mu=5.60$, $\sigma=1.27$).
A possible explanation for this support is that\,---\,regardless of whether the visualization is difficult to read\,---\,descriptions can still facilitate comprehension. For instance, SR64 noted that \hquoteB{textual description requires far less brainpower and can break down a seemingly complex visualization into an easy to grasp overview.}

% ``I've worked on different data science projects and studied statistics in college. I have the hardest time reading data visualizations when they have a tiny format or do not explain what each axis represents.''

%----------------------------------------------------------------
% \textbf{(B)~\emph{Reader Groups Disagree About Contextual Content.}}
\subsubsection{Reader Groups Disagree About Contextual Content}
\label{section:qualResultsB}
%----------------------------------------------------------------
A majority of blind readers ({63}\%, n=19) were emphatic that descriptions should \emph{not} contain an author's subjective interpretations, contextual information, or editorializing about the visualized data (i.e., Level 4 content).
Consistent with blind readers ranking this as among the least useful (Region~\ref{table:evaluation}.B), BR20 succinctly articulated a common sentiment: \hquoteA{I want the information to be simply laid out, not peppered with subjective commentary... I just prefer it to be straight facts, not presumptions or guesstimates.} BR4 also noted that an author's \hquoteA{opinions} about the data \hquoteA{should absolutely be avoided,} and BR14 emphasized {agency} when interpreting data: \hquoteA{I want to have the time and space to interpret the numbers for myself before I read the analysis.}
% \begin{displayquote}
%   \emph{\say{I don't want a description to be analytical and to include text about the possible causes of what the visualization shows. That should be separate. I want to have the time and space to interpret the numbers for myself before I read the analysis.}} (BP14)
% \end{displayquote}
% \begin{displayquote}
%   \emph{\say{I want the information to be simply laid out, not peppered with subjective commentary that may already be found in the article. While the information is useful and ties into that visualization, I just prefer it to be straight facts, not presumptions or guesstimates.
%   %I've found the most informative is laying out the numbers in a well thought out manner without being too overwhelming.
%   }} (BP20)
% \end{displayquote}
By contrast, many sighted readers {41}\% (n={37}) expressed the opposite sentiment (Region~\ref{table:evaluation}.F) noting that, for them, the most useful descriptions often \hquoteB{told a story,} communicated an important conclusion, or provided deeper insights into the visualized data.
As SR64 noted: \hquoteB{A description that simply describes the visualization and its details is hardly useful, but a description that tells a story using the data and derives a solution from it is extremely useful.}
Only {4}\% (n={4}) of sighted readers explicitly stated that a description should exclude Level 4 semantic content.

% \begin{displayquote}
%   \emph{\say{A description that simply describes the visualization and its details is hardly useful, but a description that tells a story using the data and derives a solution from it is extremely useful.
%   }} (SR64)
% \end{displayquote}

%----------------------------------------------------------------
\subsubsection{Some Readers Prefer Non-Statistical Content}
\label{section:qualResultsC}
%----------------------------------------------------------------
Overall, blind readers consistently ranked both Levels 2 and 3 as the most useful (Region~\ref{table:evaluation}.C). But, some readers explicitly expressed preference for the latter over the former, highlighting two distinguishing characteristics of Level 3 content: that it conveys not only descriptive statistics but overall perceptible trends, and that it is articulated in commonplace or ``natural''-sounding language.
For instance, BR26 remarked that a visualization description is \hquoteA{more useful if it contains the summary of the overall trends and distributions of the data rather than just mentioning some of the extreme values or means.}
Similarly, BR21 noted that \hquoteA{not everyone who encounters a data visualization needs it for statistical purposes,} and further exclaimed \hquoteA{I want to know how a layperson sees it, not a statistician; I identify more with simpler terminology.}
% Regarding the second characteristic, BR21 said the following to say.
% \begin{displayquote}
%   \emph{\say{Not everyone who encounters a data visualization needs it for statistical purposes... I want to know how a layperson sees it, not a statistician. I identify more with simpler terminology.}} (BR21)
% \end{displayquote}
These preferences help to further delineate Level 3 from Levels 2 and 4.
Content at Level 3 is ``non-statistical'' in the sense that it does only report statistical concepts and relations (as in Level 2), but neither does it do away with statistical ``objectivity'' entirely, so as to include subjective interpretation or speculation (as content in Level 4 might).
In short, Level 3 content conveys statistically-grounded concepts in not-purely-statistical terms, a challenge that is core to visualization, and science communication more broadly.

% \begin{displayquote}
%   \emph{\say{
%     I think a textual description is more useful if it contains the summary of the overall trends and distributions of the data rather than just mentioning some of the extreme values or means.}} (BR26)
% \end{displayquote}

%----------------------------------------------------------------
\subsubsection{Combinations of Content Levels Are Likely Most Useful}
\label{section:qualResultsD}
%----------------------------------------------------------------
While roughly {12}\% readers from both blind and sighted groups indicated that a description should be as concise as possible, among blind readers, {40}\% (n={12}) noted that the most useful descriptions would combine content from multiple levels.
% "shorter is better": blind n=4/30, sighted n=11/90
This finding helps to explain the bimodality in Level 1 rankings we identified in the previous section. 
According to BR9, Level 1 content is only useful if other information is also conveyed: \hquoteA{All of the descriptions provided in this survey which *only* elaborated on x/y and color-coding are almost useless.}
% \begin{displayquote}
%   \emph{\say{Descriptions of what is represented on the axes is only useful if other information is conveyed. All of the descriptions provided in this survey which *only* elaborated on x/y and color-coding are almost useless.
%   }} (BR9)
% \end{displayquote}
This sentiment was echoed by BR5, who added that if Level 1 content were \hquoteA{combined with the [Level 2 or Level 3], that'd make for a great description.}
% \begin{displayquote}
%   \emph{\say{I’d liked most of the descriptions, actually, if they were ``combined.'' For instance, some of the descriptions (I mostly rated as [least useful]) only mentioned the title and labels of the axes. If combined with the [Levels 2 or Level 3], that'd make for a great description.
%   }} (BR5)
% \end{displayquote}
% \noindent
This finding has implications for research on automatic visualization captioning: these methods should aim to generate not only the lower levels of semantic content, but to more richly communicate a chart’s overall trends and statistics, sensitive to reader preferences.

\subsubsection{Some Automatic Methods Raise Ethical Concerns}
\label{section:qualResultsE}
Research on automatically generating visualization captions is often motivated by the goal of improving information access for people with visual disabilities~\cite{qian_generating_2021, demir_summarizing_2012, obeid_chart--text_2020, qian_formative_2020}.
However, when deployed in real-world contexts, these methods may not confer their intended benefits, as one blind reader in our evaluation commented.
\begin{displayquote}
% \small
    \vspace{-1mm}
    \hquoteA{A.I. attempting to convert these images is still in its infancy. Facebook and Apple auto-descriptions of general images are more of a timewaster than useful. As a practical matter, if I find an inaccessible chart or graph, I just move on.} (BP22)
    \vspace{-1mm}
\end{displayquote}
Similarly, another participant (BR26) noted that if a description were to only describe a visualization's encodings then \hquoteA{the reader wouldn't get any insight from these texts, which not only increases the readers' reading burden but also conveys no effective information about the data.}
% \begin{displayquote}
%   \emph{\say{
%     The description would be almost useless if it only describes the encoding of the visualization because the reader wouldn't get any insight from these texts, which not only increases the readers' reading burden but also conveys no effective information about the data.}} (BR26)
% \end{displayquote}
These sentiments reflect some of the ethical concerns surrounding the deployment of nascent \ac{CV} and \ac{NLP} models, which can output accurate but minimally informative content\,---\,or worse, can output erroneous content to a trusting audience~\cite{macleod_understanding_2017, obeid_chart--text_2020}.
Facebook's automatic image descriptions, for example, have been characterized by technology educator Chancey Fleet as \say{\emph{famously useless in the Blind community}} while \say{\emph{garner[ing] a ton of glowing reviews from mainstream outlets without being of much use to disabled people}}~\cite{fleet_things_2021, hanley_computer_2021}. Such concerns might be mitigated by developing and evaluating automatic methods with disabled readers, through participatory design processes~\cite{lundgard_sociotechnical_2019}.

% Chancey Fleet. "Things which garner a ton of glowing reviews from mainstream outlets without being of much use to disabled people. for instance, facebook’s auto image descriptions, much loved by sighted journos but famously useless in the blind community," January 2021.
\section{Discussion and Future Work}
\label{section:6}
Our four-level model of semantic content\,---\,and its application to evaluating the usefulness of descriptions\,---\,has practical implications for the design of accessible data representations, and theoretical implications for the relationship between visualization and natural language.

% \subsection{Implications for Visualization Design}
\subsection{Natural Language As An Interface Into Visualization}
\label{section:practicalImplications}

Divergent reader preferences for semantic content suggests that it is helpful to think of natural language\,---\,not only as an interface for constructing and exploring visualizations~\cite{srinivasan_collecting_2021,gao_datatone_2015,setlur_eviza_2016}\,---\,but also as an {interface into visualization, for \emph{understanding} the semantic content they convey}.
Under this framing, we can apply Beaudoin-Lafon's framework for evaluating interface models in terms of their descriptive, evaluative, and generative powers~\cite{beaudouin-lafon_instrumental_2000, beaudouin-lafon_designing_2004}, to bring further clarity to the practical design implications of our model.
First, our grounded theory process yielded a model with \emph{descriptive} power: it categorizes the semantic content conveyed by visualizations.
Second, our study with blind and sighted readers demonstrated our model's \emph{evaluative} power: it offered a means of comparing different levels of semantic content, thus revealing divergent preferences between these different reader groups.
Third, future work can now begin to study our model's \emph{generative} power:
% not only its technological and ethical implications for automatically generated captions (\S~\ref{section:qualResultsE}), but also its 
its implications for novel multimodal interfaces and accessible data representations.
For instance, our evaluation suggested that descriptions primarily intending to benefit sighted readers might aim to generate higher-level semantic content (\S~\ref{section:qualResultsB}), while those intending to benefit blind readers might instead focus on affording readers the option to customize and combine different content levels (\S~\ref{section:qualResultsD}), depending on their individual preferences (\S~\ref{section:qualResultsC}).
This latter path might involve automatically \ac{ARIA} tagging web-based charts to surface semantic content at Levels 1 \& 2, with human-authors conveying Level 3 content. Or, it might involve applying our model to develop and evaluate the outputs of automatic captioning systems\,---\,to probe their technological capabilities and ethical implications\,---\,in collaboration with the relevant communities (\S~\ref{section:qualResultsE}).
To facilitate this work, we have released our corpus of visualizations and labeled sentences under an open source license: 
\href{http://vis.csail.mit.edu/pubs/vis-text-model/data/}{\texttt{http://vis.csail.mit.edu/pubs/vis-text-model/data/}}.

% \subsection{Implications for Visualization Theory}
\subsection{Natural Language As Coequal With Visualization}
\label{section:theoreticalImplications}

{In closing, we turn to a discussion of our model's implications for visualization theory.}
Not only can we think of natural language as an interface into visualization (as above), but also as an interface into data itself; coequal with and complementary to visualization.
For example, some semantic content (e.g., Level 2 statistics or Level 4 explanations) may be best conveyed via language, without any reference to visual modalities~\cite{potluri_examining_2021, hearst_would_2019}, while other content (e.g., Level 3 clusters) may be uniquely suited to visual representation.
This coequal framing is not a departure from orthodox visualization theory, but rather a return to its linguistic and semiotic origins.
Indeed, at the start of his foundational \emph{Semiology of Graphics}, Jacques Bertin introduces a similar framing to formalize an idea at the heart of visualization theory: content can be conveyed not only through speaking or writing but also through the ``language'' of graphics~\cite{bertin_semiology_1983}.
While Bertin took natural language as a point of departure for formalizing a language of graphics, we have here pursued the inverse: taking visualization as occasioning a return to language.
This theoretical inversion opens avenues for future work, for which linguistic theory and semiotics are instructive~\cite{weber_towards_2019, maceachren_visual_2012, vickers_understanding_2013}.

Within the contemporary linguistic tradition, subfields like syntax, semantics, and pragmatics suggest opportunities for further analysis at each level of our model. And, since our model focuses on English sentences and canonical chart types, extensions to other languages and bespoke charts may be warranted.
Within the semiotic tradition, Christian Metz (a contemporary of Bertin's) emphasized the \emph{pluralistic} quality of graphics~\cite{campolo_signs_2020}: the semantic content conveyed by visualizations depends not only on their graphical sign-system, but also on various ``social codes'' such as education, class, expertise, and\,---\,we hasten to include\,---\,ability.
Our evaluation with blind and sighted readers (as well as work studying how charts are deployed in particular discourse contexts~\cite{hullman_visualization_2011, hullman_content_2015, lee_viral_2021, aiello_inventorizing_2020}) 
lends credence to Metz's conception of graphics as pluralistic: different readers will have different ideas about what makes visualizations meaningful (Fig.~\ref{fig:teaser}).
As a means of revealing these differences, we have here introduced a four-level model of semantic content.
We leave further elucidation of the relationship between visualization and natural language to future work.

%% if specified like this the section will be committed in review mode
\acknowledgments{
    For their valuable feedback, we thank Emilie Gossiaux, Chancey Fleet, Michael Correll, Frank Elavsky, Beth Semel, Stephanie Tuerk, Crystal Lee, and the MIT Visualization Group.
    This work was supported by  National Science Foundation GRFP-1122374 and III-1900991.
}

\bibliographystyle{abbrv}
\bibliography{vis-text}

\begin{thebibliography}{100}

\bibitem{ackland_world_2017}
P.~Ackland, S.~Resnikoff, and R.~Bourne.
\newblock World {Blindness} and {Visual} {Impairment}.
\newblock {\em Community Eye Health}, 2017.

\bibitem{adar_communicative_2020}
E.~Adar and E.~Lee.
\newblock Communicative {Visualizations} as a {Learning} {Problem}.
\newblock In {\em {TVCG}}. IEEE, 2020.

\bibitem{aiello_inventorizing_2020}
G.~Aiello.
\newblock Inventorizing, {Situating}, {Transforming}: {Social} {Semiotics}
  {And} {Data} {Visualization}.
\newblock In M.~Engebretsen and H.~Kennedy, editors, {\em Data {Visualization}
  in {Society}}. Amsterdam University Press, 2020.

\bibitem{ali_mind-dependent_2016}
K.~M. Ali.
\newblock Mind-{Dependent} {Kinds}.
\newblock In {\em Journal of {Social} {Ontology}}, 2016.

\bibitem{amar_low-level_2005}
R.~Amar, J.~Eagan, and J.~Stasko.
\newblock Low-level {Components} {Of} {Analytic} {Activity} {In} {Information}
  {Visualization}.
\newblock In {\em {INFOVIS}}. IEEE, 2005.

\bibitem{balaji_chart-text_2018}
A.~Balaji, T.~Ramanathan, and V.~Sonathi.
\newblock Chart-{Text}: {A} {Fully} {Automated} {Chart} {Image} {Descriptor}.
\newblock {\em arXiv}, 2018.

\bibitem{beaudouin-lafon_instrumental_2000}
M.~Beaudouin-Lafon.
\newblock Instrumental {Interaction}: {An} {Interaction} {Model} {For}
  {Designing} {Post}-{WIMP} {User} {Interfaces}.
\newblock In {\em {CHI}}. ACM, 2000.

\bibitem{beaudouin-lafon_designing_2004}
M.~Beaudouin-Lafon.
\newblock Designing {Interaction}, {Not} {Interfaces}.
\newblock In {\em {AVI}}. ACM, 2004.

\bibitem{benetech_making_nodate}
Benetech.
\newblock Making {Images} {Accessible}.
\newblock http://diagramcenter.org/making-images-accessible.html/.

\bibitem{bennett_its_2021}
C.~L. Bennett, C.~Gleason, M.~K. Scheuerman, J.~P. Bigham, A.~Guo, and A.~To.
\newblock ``{It}'s {Complicated}'': {Negotiating} {Accessibility} and
  ({Mis}){Representation} in {Image} {Descriptions} of {Race}, {Gender}, and
  {Disability}.
\newblock In {\em {CHI}}. ACM, 2021.

\bibitem{bergstrom_sars-cov-2_2020}
C.~T. Bergstrom.
\newblock {SARS}-{CoV}-2 {Coronavirus}, 2020.
\newblock http://ctbergstrom.com/covid19.html.

\bibitem{bertin_semiology_1983}
J.~Bertin.
\newblock {\em Semiology of {Graphics}}.
\newblock University of Wisconsin Press, 1983.

\bibitem{bigham_vizwiz_2010}
J.~P. Bigham, C.~Jayant, H.~Ji, G.~Little, A.~Miller, R.~C. Miller, R.~Miller,
  A.~Tatarowicz, B.~White, S.~White, and T.~Yeh.
\newblock {VizWiz}: {Nearly} {Real}-time {Answers} {To} {Visual} {Questions}.
\newblock In {\em {UIST}}. ACM, 2010.

\bibitem{bigham_effects_2017}
J.~P. Bigham, I.~Lin, and S.~Savage.
\newblock The {Effects} of "{Not} {Knowing} {What} {You} {Don}'t {Know}" on
  {Web} {Accessibility} for {Blind} {Web} {Users}.
\newblock In {\em {ASSETS}}. ACM, 2017.

\bibitem{borkin_beyond_2016}
M.~A. Borkin, Z.~Bylinskii, N.~W. Kim, C.~M. Bainbridge, C.~S. Yeh, D.~Borkin,
  H.~Pfister, and A.~Oliva.
\newblock Beyond {Memorability}: {Visualization} {Recognition} and {Recall}.
\newblock In {\em {TVCG}}. IEEE, 2016.

\bibitem{borkin_what_2013}
M.~A. Borkin, A.~A. Vo, Z.~Bylinskii, P.~Isola, S.~Sunkavalli, A.~Oliva, and
  H.~Pfister.
\newblock What {Makes} a {Visualization} {Memorable}?
\newblock In {\em {TVCG}}. IEEE, 2013.

\bibitem{brehmer_multi-level_2013}
M.~Brehmer and T.~Munzner.
\newblock A {Multi}-{Level} {Typology} of {Abstract} {Visualization} {Tasks}.
\newblock In {\em {TVCG}}. IEEE, 2013.

\bibitem{campolo_signs_2020}
A.~Campolo.
\newblock Signs and {Sight}: {Jacques} {Bertin} and the {Visual} {Language} of
  {Structuralism}.
\newblock {\em Grey Room}, 2020.

\bibitem{cesal_writing_2020}
A.~Cesal.
\newblock Writing {Alt} {Text} for {Data} {Visualization}, Aug. 2020.

\bibitem{chang_defining_2009}
R.~Chang, C.~Ziemkiewicz, T.~M. Green, and W.~Ribarsky.
\newblock Defining {Insight} for {Visual} {Analytics}.
\newblock In {\em {CG}\&{A}}. IEEE, 2009.

\bibitem{chaparro_applications_2017}
A.~Chaparro and M.~Chaparro.
\newblock Applications of {Color} in {Design} for {Color}-{Deficient} {Users}.
\newblock {\em Ergonomics in Design}, 2017.

\bibitem{chen_neural_2019}
C.~Chen, R.~Zhang, S.~Kim, S.~Cohen, T.~Yu, R.~Rossi, and R.~Bunescu.
\newblock Neural {Caption} {Generation} {Over} {Figures}.
\newblock In {\em {UbiComp}/{ISWC} '19 {Adjunct}}. ACM, 2019.

\bibitem{chen_figure_2020}
C.~Chen, R.~Zhang, E.~Koh, S.~Kim, S.~Cohen, and R.~Rossi.
\newblock Figure {Captioning} with {Relation} {Maps} for {Reasoning}.
\newblock In {\em {WACV}}. IEEE, 2020.

\bibitem{chen_figure_2019}
C.~Chen, R.~Zhang, E.~Koh, S.~Kim, S.~Cohen, T.~Yu, R.~Rossi, and R.~Bunescu.
\newblock Figure {Captioning} with {Reasoning} and {Sequence}-{Level}
  {Training}.
\newblock {\em arXiv}, 2019.

\bibitem{choi_visualizing_2019}
J.~Choi, S.~Jung, D.~G. Park, J.~Choo, and N.~Elmqvist.
\newblock Visualizing for the {Non}-{Visual}: {Enabling} the {Visually}
  {Impaired} to {Use} {Visualization}.
\newblock In {\em {CGF}}. Eurographics, 2019.

\bibitem{demir_generating_2008}
S.~Demir, S.~Carberry, and K.~F. McCoy.
\newblock Generating textual summaries of bar {chartsGenerating} {Textual}
  {Summaries} {Of} {Bar} {Charts}.
\newblock In {\em {INLG}}. ACL, 2008.

\bibitem{demir_summarizing_2012}
S.~Demir, S.~Carberry, and K.~F. McCoy.
\newblock Summarizing {Information} {Graphics} {Textually}.
\newblock In {\em Computational {Linguistics}}. ACL, 2012.

\bibitem{ehrenkranz_vital_2020}
M.~Ehrenkranz.
\newblock Vital {Coronavirus} {Information} {Is} {Failing} the {Blind} and
  {Visually} {Impaired}.
\newblock {\em Vice}, 2020.

\bibitem{elavsky_chartability_2021}
F.~Elavsky.
\newblock Chartability, 2021.
\newblock https://chartability.fizz.studio/.

\bibitem{elzer_exploring_2005}
S.~Elzer, S.~Carberry, D.~Chester, S.~Demir, N.~Green, I.~Zukerman, and
  K.~Trnka.
\newblock Exploring {And} {Exploiting} {The} {Limited} {Utility} {Of}
  {Captions} {In} {Recognizing} {Intention} {In} {Information} {Graphics}.
\newblock In {\em {ACL}}. Association for Computational Linguistics, 2005.

\bibitem{elzer_browser_2007}
S.~Elzer, E.~Schwartz, S.~Carberry, D.~Chester, S.~Demir, and P.~Wu.
\newblock A {Browser} {Extension} {For} {Providing} {Visually} {Impaired}
  {Users} {Access} {To} {The} {Content} {Of} {Bar} {Charts} {On} {The} {Web}.
\newblock In {\em {WEBIST}}. SciTePress, 2007.

\bibitem{fisher_creating_2019}
C.~Fisher.
\newblock Creating {Accessible} {SVGs}, 2019.

\bibitem{fleet_things_2021}
C.~Fleet.
\newblock Things which garner a ton of glowing reviews from mainstream outlets
  without being of much use to disabled people. {For} instance, {Facebook}'s
  auto image descriptions, much loved by sighted journos but famously useless
  in the {Blind} community.
\newblock {\em Twitter}, 2021.
\newblock https://twitter.com/ChanceyFleet/status/1349211417744961536.

\bibitem{fossheim_introduction_2020}
S.~L. Fossheim.
\newblock An {Introduction} {To} {Accessible} {Data} {Visualizations} {With}
  {D3}.js, 2020.

\bibitem{galesic_graph_2011}
M.~Galesic and R.~Garcia-Retamero.
\newblock Graph {Literacy}: {A} {Cross}-cultural {Comparison}.
\newblock In {\em Medical {Decision} {Making}}. Society for Medical Decision
  Making, 2011.

\bibitem{gao_datatone_2015}
T.~Gao, M.~Dontcheva, E.~Adar, Z.~Liu, and K.~G. Karahalios.
\newblock {DataTone}: {Managing} {Ambiguity} in {Natural} {Language}
  {Interfaces} for {Data} {Visualization}.
\newblock In {\em {UIST}}. ACM, 2015.

\bibitem{garcia_advanced_2010}
S.~García, A.~Fernández, J.~Luengo, and F.~Herrera.
\newblock Advanced {Nonparametric} {Tests} {For} {Multiple} {Comparisons} {In}
  {The} {Design} {Of} {Experiments} {In} {Computational} {Intelligence} {And}
  {Data} {Mining}: {Experimental} {Analysis} {Of} {Power}.
\newblock {\em Information Sciences}, 2010.

\bibitem{geveci_vtk_2012}
B.~Geveci, W.~Schroeder, A.~Brown, and G.~Wilson.
\newblock {VTK}.
\newblock {\em The Architecture of Open Source Applications}, 2012.

\bibitem{gould_effective_2008}
B.~Gould, T.~O’Connell, and G.~Freed.
\newblock Effective {Practices} for {Description} of {Science} {Content} within
  {Digital} {Talking} {Books}.
\newblock Technical report, The WGBH National Center for Accessible Media,
  2008.
\newblock
  https://www.wgbh.org/foundation/ncam/guidelines/effective-practices-for-description-of-science-content-within-digital-talking-books.

\bibitem{hanley_computer_2021}
M.~Hanley, S.~Barocas, K.~Levy, S.~Azenkot, and H.~Nissenbaum.
\newblock Computer {Vision} and {Conflicting} {Values}: {Describing} {People}
  with {Automated} {Alt} {Text}.
\newblock {\em arXiv}, 2021.

\bibitem{hasty_guidelines_2011}
L.~Hasty, J.~Milbury, I.~Miller, A.~O'Day, P.~Acquinas, and D.~Spence.
\newblock Guidelines and {Standards} for {Tactile} {Graphics}.
\newblock Technical report, Braille Authority of North America, 2011.
\newblock http://www.brailleauthority.org/tg/.

\bibitem{hearst_would_2019}
M.~Hearst and M.~Tory.
\newblock Would {You} {Like} {A} {Chart} {With} {That}? {Incorporating}
  {Visualizations} into {Conversational} {Interfaces}.
\newblock In {\em {VIS}}. IEEE, 2019.

\bibitem{hearst_toward_2019}
M.~Hearst, M.~Tory, and V.~Setlur.
\newblock Toward {Interface} {Defaults} for {Vague} {Modifiers} in {Natural}
  {Language} {Interfaces} for {Visual} {Analysis}.
\newblock In {\em {VIS}}. IEEE, 2019.

\bibitem{hullman_visualization_2011}
J.~Hullman and N.~Diakopoulos.
\newblock Visualization {Rhetoric}: {Framing} {Effects} in {Narrative}
  {Visualization}.
\newblock In {\em {TVCG}}. IEEE, 2011.

\bibitem{hullman_contextifier_2013}
J.~Hullman, N.~Diakopoulos, and E.~Adar.
\newblock Contextifier: automatic generation of annotated stock visualizations.
\newblock In {\em {CHI}}. ACM, 2013.

\bibitem{hullman_content_2015}
J.~Hullman, N.~Diakopoulos, E.~Momeni, and E.~Adar.
\newblock Content, {Context}, and {Critique}: {Commenting} on a {Data}
  {Visualization} {Blog}.
\newblock In {\em {CSCW}}. ACM, 2015.

\bibitem{kahou_figureqa_2018}
S.~E. Kahou, V.~Michalski, A.~Atkinson, A.~Kadar, A.~Trischler, and Y.~Bengio.
\newblock {FigureQA}: {An} {Annotated} {Figure} {Dataset} for {Visual}
  {Reasoning}.
\newblock {\em arXiv}, 2018.

\bibitem{karpathy_deep_2017}
A.~Karpathy and L.~Fei-Fei.
\newblock Deep {Visual}-{Semantic} {Alignments} for {Generating} {Image}
  {Descriptions}.
\newblock In {\em {TPAMI}}. IEEE, Apr. 2017.

\bibitem{keim_literature_2007}
D.~A. Keim and D.~Oelke.
\newblock Literature {Fingerprinting}: {A} {New} {Method} for {Visual}
  {Literary} {Analysis}.
\newblock In {\em {VAST}}. IEEE, 2007.

\bibitem{kim_answering_2020}
D.~H. Kim, E.~Hoque, and M.~Agrawala.
\newblock Answering {Questions} about {Charts} and {Generating} {Visual}
  {Explanations}.
\newblock In {\em {CHI}}. ACM, Apr. 2020.

\bibitem{kim_facilitating_2018}
D.~H. Kim, E.~Hoque, J.~Kim, and M.~Agrawala.
\newblock Facilitating {Document} {Reading} by {Linking} {Text} and {Tables}.
\newblock In {\em {UIST}}. ACM, 2018.

\bibitem{kim_towards_2021}
D.~H. Kim, V.~Setlur, and M.~Agrawala.
\newblock Towards {Understanding} {How} {Readers} {Integrate} {Charts} and
  {Captions}: {A} {Case} {Study} with {Line} {Charts}.
\newblock In {\em {CHI}}. ACM, 2021.

\bibitem{kim_accessible_2021}
N.~W. Kim, S.~C. Joyner, A.~Riegelhuth, and Y.~Kim.
\newblock Accessible {Visualization}: {Design} {Space}, {Opportunities}, and
  {Challenges}.
\newblock In {\em {CGF}}. Eurographics, 2021.

\bibitem{kong_frames_2018}
H.-K. Kong, Z.~Liu, and K.~Karahalios.
\newblock Frames and {Slants} in {Titles} of {Visualizations} on
  {Controversial} {Topics}.
\newblock In {\em {CHI}}. ACM, Apr. 2018.

\bibitem{kong_trust_2019}
H.-K. Kong, Z.~Liu, and K.~Karahalios.
\newblock Trust and {Recall} of {Information} across {Varying} {Degrees} of
  {Title}-{Visualization} {Misalignment}.
\newblock In {\em {CHI}}. ACM, May 2019.

\bibitem{kong_extracting_2014}
N.~Kong, M.~A. Hearst, and M.~Agrawala.
\newblock Extracting {References} {Between} {Text} {And} {Charts} {Via}
  {Crowdsourcing}.
\newblock In {\em {CHI}}. ACM, 2014.

\bibitem{kosslyn_understanding_1989}
S.~M. Kosslyn.
\newblock Understanding {Charts} and {Graphs}.
\newblock {\em Applied Cognitive Psychology}, 1989.

\bibitem{krishna_visual_2017}
R.~Krishna, Y.~Zhu, O.~Groth, J.~Johnson, K.~Hata, J.~Kravitz, S.~Chen,
  Y.~Kalantidis, L.-J. Li, D.~A. Shamma, M.~S. Bernstein, and L.~Fei-Fei.
\newblock Visual {Genome}: {Connecting} {Language} and {Vision} {Using}
  {Crowdsourced} {Dense} {Image} {Annotations}.
\newblock In {\em {IJCV}}. Springer, 2017.

\bibitem{lai_automatic_2020}
C.~Lai, Z.~Lin, R.~Jiang, Y.~Han, C.~Liu, and X.~Yuan.
\newblock Automatic {Annotation} {Synchronizing} with {Textual} {Description}
  for {Visualization}.
\newblock In {\em {CHI}}. ACM, 2020.

\bibitem{law_characterizing_2020}
P.-M. Law, A.~Endert, and J.~Stasko.
\newblock Characterizing {Automated} {Data} {Insights}.
\newblock {\em arXiv}, 2020.

\bibitem{law_what_2020}
P.-M. Law, A.~Endert, and J.~Stasko.
\newblock What are {Data} {Insights} to {Professional} {Visualization} {Users}?
\newblock {\em arXiv}, Aug. 2020.

\bibitem{lee_viral_2021}
C.~Lee, T.~Yang, G.~Inchoco, G.~M. Jones, and A.~Satyanarayan.
\newblock Viral {Visualizations}: {How} {Coronavirus} {Skeptics} {Use}
  {Orthodox} {Data} {Practices} to {Promote} {Unorthodox} {Science} {Online}.
\newblock In {\em {CHI}}. ACM, 2021.

\bibitem{lee_vlat_2016}
S.~Lee, S.-H. Kim, and B.~C. Kwon.
\newblock Vlat: {Development} {Of} {A} {Visualization} {Literacy} {Assessment}
  {Test}.
\newblock In {\em {TVCG}}. IEEE, 2016.

\bibitem{lin_microsoft_2014}
T.-Y. Lin, M.~Maire, S.~Belongie, J.~Hays, P.~Perona, D.~Ramanan, P.~Dollár,
  and C.~L. Zitnick.
\newblock Microsoft {COCO}: {Common} {Objects} in {Context}.
\newblock In {\em {ECCV}}. Springer, 2014.

\bibitem{littlefield_covid-19_2020}
T.~Littlefield.
\newblock {COVID}-19 {Statistics} {Tracker}, 2020.
\newblock https://cvstats.net.

\bibitem{livingston_position_2020}
M.~A. Livingston and D.~Brock.
\newblock Position: {Visual} {Sentences}: {Definitions} and {Applications}.
\newblock In {\em {VIS}}. IEEE, 2020.

\bibitem{lundgard_sociotechnical_2019}
A.~Lundgard, C.~Lee, and A.~Satyanarayan.
\newblock Sociotechnical {Considerations} for {Accessible} {Visualization}
  {Design}.
\newblock In {\em {VIS}}. IEEE, Oct. 2019.

\bibitem{maceachren_visual_2012}
A.~M. MacEachren, R.~E. Roth, J.~O'Brien, B.~Li, D.~Swingley, and M.~Gahegan.
\newblock Visual {Semiotics} {Uncertainty} {Visualization}: {An} {Empirical}
  {Study}.
\newblock In {\em {TVCG}}. IEEE, 2012.

\bibitem{macleod_understanding_2017}
H.~MacLeod, C.~L. Bennett, M.~R. Morris, and E.~Cutrell.
\newblock Understanding {Blind} {People}'s {Experiences} with
  {Computer}-{Generated} {Captions} of {Social} {Media} {Images}.
\newblock In {\em {CHI}}. ACM, 2017.

\bibitem{matejka_same_2017}
J.~Matejka and G.~Fitzmaurice.
\newblock Same {Stats}, {Different} {Graphs}: {Generating} {Datasets} with
  {Varied} {Appearance} and {Identical} {Statistics} through {Simulated}
  {Annealing}.
\newblock In {\em {CHI}}. ACM, 2017.

\bibitem{moraes_evaluating_2014}
P.~Moraes, G.~Sina, K.~McCoy, and S.~Carberry.
\newblock Evaluating {The} {Accessibility} {Of} {Line} {Graphs} {Through}
  {Textual} {Summaries} {For} {Visually} {Impaired} {Users}.
\newblock In {\em {ASSETS}}. ACM, 2014.

\bibitem{morash_guiding_2015}
V.~S. Morash, Y.-T. Siu, J.~A. Miele, L.~Hasty, and S.~Landau.
\newblock Guiding {Novice} {Web} {Workers} in {Making} {Image} {Descriptions}
  {Using} {Templates}.
\newblock In {\em {TACCESS}}. ACM, 2015.

\bibitem{morris_rich_2018}
M.~R. Morris, J.~Johnson, C.~L. Bennett, and E.~Cutrell.
\newblock Rich {Representations} of {Visual} {Content} for {Screen} {Reader}
  {Users}.
\newblock In {\em {CHI}}. ACM, 2018.

\bibitem{olson_curiosity_2014}
M.~Muller.
\newblock Curiosity, {Creativity}, and {Surprise} as {Analytic} {Tools}:
  {Grounded} {Theory} {Method}.
\newblock In J.~S. Olson and W.~A. Kellogg, editors, {\em Ways of {Knowing} in
  {HCI}}. Springer, 2014.

\bibitem{narechania_nl4dv_2021}
A.~Narechania, A.~Srinivasan, and J.~Stasko.
\newblock {NL4DV}: {A} {Toolkit} for {Generating} {Analytic} {Specifications}
  for {Data} {Visualization} from {Natural} {Language} {Queries}.
\newblock In {\em {TVCG}}. IEEE, 2021.

\bibitem{north_toward_2006}
C.~North.
\newblock Toward {Measuring} {Visualization} {Insight}.
\newblock In {\em {CG}\&{A}}. IEEE, 2006.

\bibitem{nunez_optimizing_2018}
J.~R. Nuñez, C.~R. Anderton, and R.~S. Renslow.
\newblock Optimizing {Colormaps} {With} {Consideration} {For} {Color} {Vision}
  {Deficiency} {To} {Enable} {Accurate} {Interpretation} {Of} {Scientific}
  {Data}.
\newblock {\em PLOS ONE}, 2018.

\bibitem{obeid_chart--text_2020}
J.~Obeid and E.~Hoque.
\newblock Chart-to-{Text}: {Generating} {Natural} {Language} {Descriptions} for
  {Charts} by {Adapting} the {Transformer} {Model}.
\newblock {\em arXiv}, 2020.

\bibitem{oliveira_towards_2013}
M.~M. Oliveira.
\newblock Towards {More} {Accessible} {Visualizations} for
  {Color}-{Vision}-{Deficient} {Individuals}.
\newblock In {\em {CiSE}}. IEEE, 2013.

\bibitem{ottley_curious_2019}
A.~Ottley, A.~Kaszowska, R.~J. Crouser, and E.~M. Peck.
\newblock The {Curious} {Case} of {Combining} {Text} and {Visualization}.
\newblock In {\em {EuroVis}}. Eurographics, 2019.

\bibitem{poco_reverse-engineering_2017}
J.~Poco and J.~Heer.
\newblock Reverse-{Engineering} {Visualizations}: {Recovering} {Visual}
  {Encodings} from {Chart} {Images}.
\newblock In {\em {CGF}}. Eurographics, 2017.

\bibitem{potluri_examining_2021}
V.~Potluri, T.~E. Grindeland, J.~E. Froehlich, and J.~Mankoff.
\newblock Examining {Visual} {Semantic} {Understanding} in {Blind} and
  {Low}-{Vision} {Technology} {Users}.
\newblock In {\em {CHI}}. ACM, 2021.

\bibitem{qian_formative_2020}
X.~Qian, E.~Koh, F.~Du, S.~Kim, and J.~Chan.
\newblock A {Formative} {Study} on {Designing} {Accurate} and {Natural}
  {Figure} {Captioning} {Systems}.
\newblock In {\em {CHI} {EA}}. ACM, 2020.

\bibitem{qian_generating_2021}
X.~Qian, E.~Koh, F.~Du, S.~Kim, J.~Chan, R.~A. Rossi, S.~Malik, and T.~Y. Lee.
\newblock Generating {Accurate} {Caption} {Units} for {Figure} {Captioning}.
\newblock In {\em {WWW}}. ACM, 2021.

\bibitem{royer_developing_2001}
J.~M. Royer.
\newblock Developing {Reading} {And} {Listening} {Comprehension} {Tests}
  {Based} {On} {The} {Sentence} {Verification} {Technique} ({SVT}).
\newblock In {\em Journal of {Adolescent} \& {Adult} {Literacy}}. International
  Literacy Association, 2001.

\bibitem{royer_sentence_1979}
J.~M. Royer, C.~N. Hastings, and C.~Hook.
\newblock A {Sentence} {Verification} {Technique} {For} {Measuring} {Reading}
  {Comprehension}.
\newblock {\em Journal of Reading Behavior}, 1979.

\bibitem{satyanarayan_vega-lite_2017}
A.~Satyanarayan, D.~Moritz, K.~Wongsuphasawat, and J.~Heer.
\newblock Vega-{Lite}: {A} {Grammar} of {Interactive} {Graphics}.
\newblock In {\em {TVCG}}. IEEE, 2017.

\bibitem{schepers_why_2020}
D.~Schepers.
\newblock Why {Accessibility} {Is} at the {Heart} of {Data} {Visualization},
  2020.

\bibitem{setlur_eviza_2016}
V.~Setlur, S.~E. Battersby, M.~Tory, R.~Gossweiler, and A.~X. Chang.
\newblock Eviza: {A} {Natural} {Language} {Interface} for {Visual} {Analysis}.
\newblock In {\em {UIST}}. ACM, 2016.

\bibitem{setlur_inferencing_2019}
V.~Setlur, M.~Tory, and A.~Djalali.
\newblock Inferencing {Underspecified} {Natural} {Language} {Utterances} {In}
  {Visual} {Analysis}.
\newblock In {\em {IUI}}. ACM, 2019.

\bibitem{sharif_understanding_2021}
A.~Sharif, S.~S. Chintalapati, J.~O. Wobbrock, and K.~Reinecke.
\newblock Understanding {Screen}-{Reader} {Users}’ {Experiences} with
  {Online} {Data} {Visualizations}.
\newblock In {\em {ASSETS}}. ACM, 2021.

\bibitem{srinivasan_augmenting_2019}
A.~Srinivasan, S.~M. Drucker, A.~Endert, and J.~Stasko.
\newblock Augmenting {Visualizations} with {Interactive} {Data} {Facts} to
  {Facilitate} {Interpretation} and {Communication}.
\newblock In {\em {TVCG}}. IEEE, 2019.

\bibitem{srinivasan_collecting_2021}
A.~Srinivasan, N.~Nyapathy, B.~Lee, S.~M. Drucker, and J.~Stasko.
\newblock Collecting and {Characterizing} {Natural} {Language} {Utterances} for
  {Specifying} {Data} {Visualizations}.
\newblock In {\em {CHI}}. ACM, 2021.

\bibitem{sutton_accessible_2020}
H.~Sutton.
\newblock Accessible {Covid}-19 {Tracker} {Enables} {A} {Way} {For} {Visually}
  {Impaired} {To} {Stay} {Up} {To} {Date}.
\newblock {\em Disability Compliance for Higher Education}, 2020.

\bibitem{tang_extracting_2017}
B.~Tang, S.~Han, M.~L. Yiu, R.~Ding, and D.~Zhang.
\newblock Extracting {Top}-{K} {Insights} from {Multi}-dimensional {Data}.
\newblock In {\em {SIGMOD}}. ACM, 2017.

\bibitem{team_bokeh_2014}
B.~D. Team.
\newblock {\em Bokeh: {Python} {Library} {For} {Interactive} {Visualization}}.
\newblock Bokeh Development Team, 2014.

\bibitem{vickers_understanding_2013}
P.~Vickers, J.~Faith, and N.~Rossiter.
\newblock Understanding {Visualization}: {A} {Formal} {Approach} {Using}
  {Category} {Theory} and {Semiotics}.
\newblock In {\em {TVCG}}. IEEE, 2013.

\bibitem{vinyals_show_2015}
O.~Vinyals, A.~Toshev, S.~Bengio, and D.~Erhan.
\newblock Show and {Tell}: {A} {Neural} {Image} {Caption} {Generator}.
\newblock In {\em {CVPR}}, 2015.

\bibitem{w3c_wai_2019}
W3C.
\newblock {WAI} {Web} {Accessibility} {Tutorials}: {Complex} {Images}, 2019.
\newblock https://www.w3.org/WAI/tutorials/images/complex/.

\bibitem{wang_datashot_2020}
Y.~Wang, Z.~Sun, H.~Zhang, W.~Cui, K.~Xu, X.~Ma, and D.~Zhang.
\newblock {DataShot}: {Automatic} {Generation} of {Fact} {Sheets} from
  {Tabular} {Data}.
\newblock In {\em {TVCG}}. IEEE, 2020.

\bibitem{watson_accessible_2017-1}
L.~Watson.
\newblock Accessible {SVG} {Line} {Graphs}, 2017.
\newblock https://tink.uk/accessible-svg-line-graphs/.

\bibitem{watson_accessible_2018}
L.~Watson.
\newblock Accessible {SVG} {Flowcharts}, 2018.

\bibitem{weber_towards_2019}
W.~Weber.
\newblock Towards a {Semiotics} of {Data} {Visualization} – an {Inventory} of
  {Graphic} {Resources}.
\newblock In {\em {IV}}. IEEE, 2019.

\bibitem{wilkinson_grammar_2005}
L.~Wilkinson.
\newblock {\em The {Grammar} of {Graphics}}.
\newblock Statistics and {Computing}. Springer-Verlag, 2005.

\bibitem{wu_understanding_2021}
K.~Wu, E.~Petersen, T.~Ahmad, D.~Burlinson, S.~Tanis, and D.~A. Szafir.
\newblock Understanding {Data} {Accessibility} for {People} with {Intellectual}
  and {Developmental} {Disabilities}.
\newblock In {\em {CHI} 2021}, 2021.

\bibitem{xiong_curse_2020}
C.~Xiong, L.~V. Weelden, and S.~Franconeri.
\newblock The {Curse} of {Knowledge} in {Visual} {Data} {Communication}.
\newblock In {\em {TVCG}}. IEEE, 2020.

\bibitem{xu_show_2016}
K.~Xu, J.~Ba, R.~Kiros, K.~Cho, A.~Courville, R.~Salakhutdinov, R.~Zemel, and
  Y.~Bengio.
\newblock Show, {Attend} and {Tell}: {Neural} {Image} {Caption} {Generation}
  with {Visual} {Attention}.
\newblock {\em arXiv}, 2016.

\bibitem{yi_understanding_2008}
J.~S. Yi, Y.-a. Kang, J.~T. Stasko, and J.~A. Jacko.
\newblock Understanding and {Characterizing} {Insights}: {How} {Do} {People}
  {Gain} {Insights} {Using} {Information} {Visualization}?
\newblock In {\em {BELIV}}. ACM, 2008.

\end{thebibliography}
\end{document}